\documentclass[12pt]{article}
\usepackage{amssymb,epsf,graphicx}
\newcommand{\be}{\begin{equation}}
\newcommand{\ee}{\end{equation}}
\newcommand{\bea}{\begin{eqnarray}}
\newcommand{\eea}{\end{eqnarray}}
\newcommand{\bdm}{\begin{displaymath}}
\newcommand{\edm}{\end{displaymath}}

\def\a{\alpha}
\def\b{\beta}

\def\d{\delta}
\def\e{\epsilon}           
\def\f{\phi}               
  
\def\g{\gamma}

\def\k{\kappa}             
\def\lam{\lambda}
\def\m{\mu}
\def\n{\nu}
  
\def\r{\rho}                                     

\def\G{\Gamma}

\def\pa{\partial}

\newcommand{\Tr}{\mathop{\rm Tr}\nolimits}
\newcommand{\re}{\mathop{\rm Re}\nolimits}
\newcommand{\im}{\mathop{\rm Im}\nolimits}
\newcommand{\asympt}{\mathop{\sim}}
\newcommand{\arctanh}{\mathop{\rm arctanh}\nolimits}
\newcommand{\sn}{\mathop{\rm sn}\nolimits}

\def\l{\left}                    
\def\r{\right}
\let\eps = \varepsilon

\def\W{{\bf W}}

\begin{document}
\baselineskip=15.5pt
\pagestyle{plain}
\setcounter{page}{1}

\begin{flushright}
CTP-MIT-3450\\
ITFA-2003-55 \\
{\tt hep-th/0312055}
\end{flushright}

\vskip 2cm

\begin{center}
{\Large \bf Fake Supergravity and Domain Wall Stability \\}
\vskip 1cm

{\bf D.~Z.~Freedman$^{1,2}$, C.~N\'u\~nez$^2$, M.~Schnabl$^2$, K.~Skenderis$^3$} \\
\vskip 0.5cm
{\it $^1$  Department of Mathematics, Massachusetts Institute of Technology,\\
Cambridge, MA 02139, USA,} \\
{\tt E-mail:dzf@math.mit.edu} \\
\medskip
{\it $^2$ Center for Theoretical Physics, Massachusetts Institute of Technology,\\
Cambridge, MA 02139, USA,} \\
{\tt E-mail: nunez, schnabl@lns.mit.edu}  \\
\medskip
{\it $^3$  Institute for  Theoretical Physics, University of Amsterdam,\\
 Valkenierstraat 65, 1018 XE Amsterdam, The Netherlands,} \\
{\tt E-mail: skenderi@science.uva.nl}
\end{center}

\vskip1cm

\begin{center}
{\bf Abstract}
\end{center}
\medskip

We review the generalized Witten-Nester spinor stability argument for flat
domain wall solutions of gravitational theories. Neither the field theory nor
the solution need be supersymmetric. Nor is the space-time dimension
restricted. We develop the non-trivial extension required for $AdS$-sliced
domain walls and apply this to show that the recently proposed ``Janus''
solution of Type IIB supergravity is stable non-perturbatively for a broad class of
deformations. Generalizations of this solution to arbitrary dimension and a
simple curious linear dilaton solution of Type IIB supergravity
are byproducts of this work.

\newpage

\section{Introduction}

Many domain wall solutions of supergravity theories have been
studied in the literature in order to explore the AdS/CFT
correspondence, to find a fundamental setting for
brane world cosmology, and for other reasons. In this paper we
will review and extend stability arguments for domain walls based on the
elegant spinor methods of the Witten positive energy theorem
and its generalizations \cite{Witten, Nester, GibHulWar, Boucher, Towns,Cvetic}.
\footnote {See \cite{Traschen} for a recent paper with similar
aims which discusses the stability of $p$-brane spacetimes.}

Many solutions studied in the past are supersymmetric. One would
expect these to be stable, and there are known arguments
which use the transformation rules of the supergravity theory and
the Killing spinors supported by the solutions. Yet these
arguments do not apply to the many solutions with curvature
singularities.

Non-supersymmetric solutions are also known and might well be
important since SUSY is certainly broken in our universe. Most
domain wall solutions, both SUSY and non-SUSY, are planar; the
isometry group of their metrics
\bea \label{domwall}
ds^2 &=& e^{2A(r)} \eta_{ij} dx^i dx^j + e^{2h(r)}dr^2\\
 \eta_{ij} &=& \rm{diag}(-1,1,1,....,1) \nonumber
\eea
is the Poincar\'{e} group in $d$ flat space-time dimensions. (The choice
$h(r)=0$ is convenient for many purposes but we keep $h(r)$ unfixed to
facilitate comparison with different radial coordinates used in the
literature.)

For planar domain walls there is a formal stability argument
\cite{Towns,SkenTowns} based on what we propose to call ``fake supergravity''.
In fake supergravity one defines a spinor energy using fake transformation
rules similar to those of a real supergravity theory, but containing a
superpotential $W(\phi)$ which is not that of the real theory. Instead
$W(\phi)$ must satisfy an equation which relates it to the scalar potential
$V(\phi)$, and one can formulate certain first order equations (the fake BPS
equations) whose solutions automatically satisfy the second order Einstein
field equations for domain wall metrics (\ref{domwall}) and the accompanying
scalar field $\phi(r).$ If one can find a  $W(\phi)$ such that the domain wall
solution under test is a solution of the first order system, then that domain
wall is stable, if there are no singularities.\footnote{Further conditions are
discussed in Sec. \ref{Sec-Stability-flat}.}
One curious feature of fake supergravity is that it can
work in any space-time dimension $d$, whereas real supergravity is limited to
$d \le 11.$

Domain walls with the isometry group $SO(d-1,2)$ of the space-time
$AdS_d$ have also been studied \cite{Lust,Sabra,Behrndt}. Their metrics take
the form
\be \label{adswall}
ds^2 = e^{2A(r)} g_{ij}(x) dx^i dx^j + e^{2h(r)} dr^2
\ee
where $g_{ij}(x)$ is a metric on $AdS_d$ with scale $L_d$.
A domain wall of this type was recently found \cite{Gutp} as a
solution of type IIB supergravity. The solution contains a flowing dilaton
$\phi(r)$, but no other $r$-dependent matter fields, and there is
an accompanying round $S_5$ internal space. The solution is
regular if one chooses parameters such that the rate of variation
of the dilaton is sufficiently slow.

In this paper we develop stability arguments for non-singular $AdS_d$-sliced
domain walls. A non-trivial extension of the fake supergravity approach,
related to the work of \cite{Lust} in real $D=5$ supergravity, is required for
this. This argument gives a definition of energy which vanishes for the
background solution itself and is positive for fluctuations about the
background which obey suitable boundary conditions.

These arguments imply that the solution of \cite{Gutp} enjoys non-perturbative
stability with respect to fluctuations of the metric and dilaton while other
fields of Type IIB supergravity remain fixed at their vacuum values. The
formalism we develop can accommodate additional fields, but it becomes more
difficult to establish the required properties of the superpotential. To
remediate this difficulty we work in the spirit of \cite{Boucher} and derive
inequalities which show that the ``Janus'' solution \cite{Gutp} is also stable within
several different consistent truncations of Type IIB supergravity. Some of these
truncations include negative $m^2$ fields and potentials unbounded below. These
indications of global stability make it more compelling to understand the
AdS/CFT dual of the solution proposed in \cite{Gutp}. We do not discuss this
here.

It is not guaranteed that a given solution of the field equations can be
reproduced in the framework of fake supergravity. Indeed there are known
solutions which are pure $AdS$ metrics (those with $A(r) \equiv r/L$ in
(\ref{domwall}) if $h(r)=0$ and with fixed scalars) which are unstable because
small fluctuations violate the stability bound of \cite{brfr}. In general it is
not always possible to satisfy the required conditions on the superpotential.

In Sec. 2 we discuss the equations of motion satisfied by domain
walls
and present some simple examples of non-supersymmetric domain
wall solutions of Type IIB supergravity. They involve a single
flowing scalar field, the dilaton.
These simple dilaton domain walls are the prototype solutions we study.
The fake supergravity stability argument for planar domain walls
(\ref{domwall}) is reviewed in Sec 3. In Sec 4. we extend
this argument to $AdS_d$ domain walls (\ref{adswall}).
Sec. 5 is devoted to stability arguments for the solution of
\cite{Gutp} for non-dilatonic fluctuations. In Sec 6.
we discuss a very simple and apparently new solution of Type IIB
which emerged from the techniques of Sec. 4.

\section{Domain Walls: Basics and Examples}
\setcounter{equation}{0}
\label{Section_DW}

We consider a scalar-gravity action in $d+1$ dimensions:
\be\label{toymod}
S =  \int d^{d+1}x \sqrt{-g} \l[ \frac{1}{2\kappa^2} R -
\frac{1}{2} \partial_\mu \phi \partial^\mu \phi - V(\phi)\r].
\ee
Such actions can arise via Kaluza-Klein reduction of a still
higher dimensional theory. Although we include only one scalar
explicitly, additional scalars (with $\sigma$-model interactions)
and higher rank bosonic fields can be included. The equations of
motion are
\bea \label{geneom}
\frac{1}{\k^2} R_{\m\n} &=& \partial_\m \phi \partial_\n \phi + \frac{2}{d-1}
g_{\m\n} V(\phi)
\nonumber\\
\Box \phi &=& \partial V/\partial \phi
\eea

We assume that the potential $V(\phi)$ has a critical point at
$\phi =\phi_0$, with $V_0 \equiv V(\phi_0) < 0$. Thus one solution of
(\ref{geneom}) is $AdS_{d+1}$ with scale $L$. In this case we have
\bea \label{adssol}
R_{\mu\nu} &=& - \frac{d}{L^2} g_{\mu\nu}
\nonumber\\
V_0 &=& -\frac{d(d-1)}{2L^2 \k^2}.
\eea
We will introduce explicit parameterizations of the $AdS$ metric
$g_{\mu\nu}$ when needed.

\subsection{Flat domain walls.}
We are more interested in domain wall solutions of (\ref{geneom})
with $r$-dependent scalar $\phi(r)$ and metrics of the form
(\ref{domwall}) or (\ref{adswall}), and we require that these
approach the $AdS_{d+1}$ geometry at the boundary. With the
coordinate choice $h(r)=0$, the boundary occurs as $r \rightarrow
+ \infty$. Frame and connection 1-forms and curvature tensors for
our presentation of domain walls are given in Appendix A.

We first consider flat domain walls. When the metric of (\ref{domwall}) and the
restriction $\phi = \phi(r)$ are incorporated, the Einstein and scalar
equations of motion of (\ref{geneom}) reduce to ordinary differential equations
in $r$, namely
\bea \label{dweom}
 A'' -  A'h' &=& - \frac{\kappa^2}{d-1} {\phi'}^2
\nonumber\\
{A'}^2 &=&  \frac{\kappa^2}{d(d-1)} {\phi'}^2 - \frac{2\kappa^2}{d(d-1)} V(\phi) e^{2h}
\nonumber\\
\phi'' + (d A' - h') \phi' &=& \frac{\partial V}{\partial \phi} e^{2h}
\eea

It is quite well known \cite{SkenTowns,dfgk} that any
solution of the following first order flow equations is also a solution
of (\ref{dweom}):
\be \label{flowA}
A'(r) = 2 e^h W(\phi(r))
\ee
\be \label{flowphi}
\phi'(r) = -\frac{2(d-1)}{\kappa^2} e^h \partial_{\phi} W(\phi(r)).
\ee
The superpotential $W(\phi)$ is related to the potential $V(\phi)$
by\footnote{The prime in $W'$ denotes a derivative with respect to $\phi$, 
whereas the prime attached to the fields $\phi, A, h$ denotes a derivative 
with respect to the radial coordinate $r$.}
\be \label{VWrel1}
\kappa^2 V(\phi) = 2(d-1)^2 \l( \frac{1}{\kappa^2} {W'}^2 - \frac{d}{d-1} W^2 \r).
\ee
These fake BPS equations for flat domain walls will be rederived
in the next section.

The simplest example of a domain wall is the following solution of
(\ref{dweom}) (with $h(r)=0$) for the theory with constant potential
$V(\phi) = V_0$ of (\ref{adssol}). The scalar satisfies $\phi'(r)
= c \exp(-d A(r))$. After routine integration, one finds
\bea\label{dilwall}
\phi(r) &=& \phi_0 + \sqrt{\frac{d-1}{d\, \kappa^2 }}
\log \frac{1-e^{-d (r-r^{*})/L}}{1+e^{-d (r-r^{*})/L}}
\nonumber\\
A(r) &=& A_0 + \frac{r-r^{*}}{L} +
\frac{1}{d} \log \l(1-e^{-2 d (r-r^{*})/L} \r),
\eea
where $c$ is related to the other integration constants by
$c=\frac{\sqrt{d(d-1)}}{\kappa L} e^{d A_0}$.
This gives an asymptotically $AdS$ geometry with boundary region
$r \rightarrow \infty$, but there is a curvature singularity at
$r=r^{*}$.

When $d=4$ this is the dilaton domain wall  solution of Type IIB supergravity
which was found and studied \cite{Sfetsos,Gubser,Neil} in the early period of
the $AdS/CFT$ correspondence. As a solution of IIB supergravity, it is not
supersymmetric. There are no true Killing spinors, since the dilatino condition
from the Type IIB supergravity transformation rules
\be
\delta\chi= \frac{i}{2}\gamma^\mu (\partial_\mu \phi + i e^{\phi}
\partial_\mu \xi)\epsilon^* - \frac{i}{24}\gamma^{\mu\nu\rho}(e^{-\phi/2}
H^{(NS)}_{\mu\nu\rho}  + i e^{\phi/2}
F^{(RR)}_{\mu\nu\rho})\epsilon
\ee
cannot be satisfied because the axion $\xi$ and 3-forms vanish. The indices
$\mu, \nu, \rho$ are 10-dimensional coordinate indices.

We will now show that there is a superpotential $W(\phi)$ such that
(\ref{dilwall}) is also a solution of (\ref{flowA}, \ref{flowphi},
\ref{VWrel1}) for any dimension $d$. We thus achieve fake supersymmetry, as we
will confirm by exhibiting fake Killing spinors in the next section. The
obvious constant $W= 1/2L$ does not work, but with the general solution of
(\ref{VWrel}), namely
\be
W(\phi) =  \frac{1}{2L} \cosh \l( \kappa
\sqrt{\frac{d}{d-1}}(\phi-\phi_0) \r)
\ee
one can easily integrate (\ref{flowA}, \ref{flowphi}) and find that the
solution agrees with (\ref{dilwall}). The constant $r^*$ arises as an
integration constant.

Note that we have chosen the solution of (\ref{VWrel1}) which is positive near
the boundary value $\phi \sim \phi_0$, and we have chosen signs in
(\ref{flowA}, \ref{flowphi}) so that the boundary of the geometry appears as $r
\rightarrow +\infty$. These conventions are natural for the extension to
$AdS_d$ domain walls in Sec. 4, but they differ from some earlier applications.

Let us use the term {\it adapted superpotential} to denote the
particular $W(\phi)$ for which the first order flow equations
produce a given domain wall solution of (\ref{dweom}).
For non-constant $V(\phi)$, it may not be possible to solve
(\ref{VWrel1}) and find the superpotential $W(\phi)$
explicitly. This may be inconvenient, but to establish fake
supersymmetry we need only know that the adapted superpotential
exist for a given solution $A(r),~~\phi(r)$ of (\ref{dweom}). If
$\phi(r)$ is monotonic, the inverse function $r(\phi)$ exists.
One may then use (\ref{flowA}) to {\it define} the adapted superpotential.

\subsection{$AdS_d$-sliced domain walls.}
We now discuss the equations of motion for $AdS_d$-sliced domain
walls of co-dimension one.
Frames, connections and curvatures for the metric
(\ref{adswall}) are given in Appendix A. When inserted in
(\ref{geneom}) one finds that wall profile $A(r)$ and scalar
$\phi(r)$ obey the coupled equations which are modifications of (\ref{dweom}),
\bea\label{adseom}
 A'' -  A'h' &=& - \frac{\kappa^2}{d-1} {\phi'}^2 + \frac{1}{L_d^2} e^{-2A+2h}
\nonumber\\
{A'}^2 &=&  \frac{\kappa^2}{d(d-1)} {\phi'}^2 - \frac{2\kappa^2}{d(d-1)} V(\phi) e^{2h}
- \frac{1}{L_d^2} e^{-2A+2h}
\nonumber\\
\phi'' + (d A' - h') \phi' &=& \frac{\partial V}{\partial \phi} e^{2h}
\eea

A set of first order equations which extend (\ref{flowA} -- \ref{VWrel1}) to
$AdS_d$-sliced walls was presented in \cite{dfgk}. These equations are
\bea \label{dfgkflow}
A'(r) &=& 2 \g(r)\, e^h\, W(\phi(r))
\nonumber\\
\phi'(r) &=& -\frac{1}{ \g(r)} \frac{2(d-1)}{\kappa^2} e^h \frac{\partial W}{\partial
\phi}
\nonumber\\
V(\phi) &=& \frac{2(d-1)^2}{\kappa^2}
\l( \frac{1}{\kappa^2 \g(r)^2} {W'}^2 - \frac{d}{d-1} W^2 \r)
\eea
which differ from (\ref{flowA} -- \ref{VWrel1}) by the inclusion of the factor
\be\label{gamdef}
\g(r) \equiv \sqrt{1 - \frac{e^{-2A(r)}}{4 L_d^2 W(\phi(r))^2}}.
\ee
The constant $L_d$ is the $AdS_d$ scale, and one obtains the previous
(\ref{flowA} -- \ref{VWrel1}) as $L_d \rightarrow \infty$. The
system (\ref{dfgkflow}) is well-posed \cite{dfgk}, but it is
rather unworkable. In Sec. 4 we will derive an alternate set of first order
equations which involves an $su(2)$-valued superpotential ${\bf W}(\phi) =
W_a(\phi) \tau^a$, where the $\tau^a$ are the Pauli matrices. The structure of
the new equations is even simpler than
 (\ref{flowA} -- \ref{VWrel1}) and they are easily solved,
given ${\bf W}(\phi)$. However, ${\bf W}(\phi)$ must satisfy a
nonlinear condition in addition to (\ref{VWrel1}).
We show that any solution of the new equations
also satisfies (\ref{dfgkflow}).

\subsection{Janus solutions.}

A simple example of an $AdS_d$-sliced domain wall is the extension to $d$
dimensions of the dilaton domain wall solution of Type IIB supergravity of
\cite{Gutp}. We take the constant potential $V(\phi) = V_0$, see
(\ref{adssol}), and we proceed as in \cite{Gutp}, but use the radial coordinate
$r$ for which $h(r) = 0$. We take $\phi' =c \exp(-dA(r))$ so the scalar
equation of (\ref{adseom}) is satisfied. Any solution of the wall profile
equation
\be \label{profeq}
A'^2 = (1/L^2) [1 -e^{-2A} + b e^{-2dA}],
\ee
will also satisfy the equation involving $A''$ in
(\ref{adseom}).
The constant $b$ is related to $c$ by $b=\frac{\kappa^2}{d(d-1)} c^2L^2$.
We have set $L_d=L$ for simplicity.

When $b=0$, the solution gives pure $AdS_{d+1}$ in the form
\be
ds^2 = \cosh^2(r/L) g_{ij}(x) dx^i dx^j  +  dr^2.
\ee

For $b \ne 0$ we will not be able to solve (\ref{profeq})
exactly (unless $d=2$), and we need the following argument similar to that of
\cite{Gutp}. With $x \equiv e^{-2A}$, we consider the polynomial
$P(x) \equiv b x^d -x +1.$ For small b, there are exactly two
real zeros (which occur for $x>1$). This continues to be true for
\be
0 < b < b_0 \equiv \frac{1}{d} \l(\frac{d-1}{d}\r)^{d-1}.
\ee
At $b=b_0$ the zeros coalesce at $x_0 = (b_0 d)^{-\frac{1}{d-1}}$ and
become complex for $b>b_0$.

This behavior is relevant to the physics, as we
can see from the implicit solution of (\ref{profeq}), namely
\be \label{stuff}
r = \int_{A_0}^A \frac{dA}{\sqrt{1 -e^{-2A} + b e^{-2dA}}}.
\ee
The lower limit $A_0$ will be specified below.
For $b>b_0$, there is no natural lower bound on the variable $A$
and the geometry would be geodesically incomplete unless extended
to $A \rightarrow -\infty$ where there is a curvature singularity.

Therefore we restrict to the
range $0 < b < b_0$ in which the minimum value of $A_{min}$ is
given by $A_{min} = - \ln(x_{min})/2$, where $x_{min}$ is the smallest zero of
$P(x)$. The formula (\ref{stuff}), with $A_0=A_{min}$ thus
defines half the geometry,
namely the region $0 \le r < +\infty,~~ A_{min} <A(r)<+\infty$.
This $r>0$ region is not geodesically complete.
But all odd order derivatives of $A(r)$ vanish at $r=0$, so that $A(r)$
can be extended to the region $ -\infty < r <0$ as even function, $A(r)=
A(-r)$, and the continued function is $C^{\infty}$.
The full geometry is geodesically complete and has two boundary
regions, namely $r \rightarrow \pm \infty.$

With $A(r)$ defined above, the dilaton is given by
\be
\phi(r) = \phi_0 + c \int_0^r e^{-dA(r)} dr.
\ee
It is monotonic, and odd in $r$ except for the additive integration
constant $\phi_0$. In the boundary regions $r \rightarrow \pm \infty$,
it approaches the limits
\be
\phi(r) \rightarrow \phi_{\pm} \equiv  \phi_0 \pm c \int_0^{\infty}
e^{-dA(r)} dr.
\ee

Our choice of radial coordinate $r$ (with $h(r)=0$) was motivated by the fact
that (\ref{stuff}) can be integrated in terms of elementary functions when d=2.
This leads to the explicit presentation of the $d=2$ solution discussed in
Appendix \ref{App-d2}.  However, the space-time geometry is most easily
visualized using a radial coordinate of finite range. Therefore, in the rest of
this section we switch to the notation of \cite{Gutp}, with radial variable
$\mu$ (corresponding to the case $h =A$ in the notation above), and we use a
standard global metric on the $AdS_d$ slices.

The metric (\ref{adswall}) then takes the form
\be \label{globmet}
ds^2 = \frac{L^2 e^{2A(\mu)}}{\cos^2\lam} \l[-dt^2 +\cos^2\lam \, d\mu^2 +
d\lam^2 + \rm{sin}^2\lam \, d\Omega^2_{d-2} \r]
\ee
where the range of the principal coordinate of $AdS_d$ is $0\le
\lam < \frac{\pi}{2}$ for $d>2$, but $-\frac{\pi}{2}
<\lam<\frac{\pi}{2}$ when $d=2$, and $d\Omega^2_{d-2}$ is a
metric on the unit sphere $S_{d-2}$. The wall profile $A(\mu)$ is
defined implicitly by the integral
\be \label{muint}
\mu = \int_{A_0}^A \frac{dA}{\sqrt{e^{2A} -1 + b e^{-2(d-1)A}}}.
\ee
It can be extended to negative $\mu$ as discussed above. The
range of $\mu$ is $-\mu_0 <\mu < \mu_0$. The boundary limit $\mu_0$ can, in
principle, be obtained from the integral (\ref{muint}), with
upper limit $A \rightarrow +\infty.$ For small $b$ and general
$d$, one finds the series expansion
\be\label{mu0exp}
\mu_{0} = \frac{\pi}{2} \l( 1 +
\frac{\Gamma\!\l(d+\frac{1}{2}\r)}{\Gamma(d)\Gamma\!\l(\frac{1}{2}\r)}b +
\frac{\Gamma\!\l(2d+\frac{1}{2}\r)}{\Gamma(2d-1)\Gamma\!\l(\frac{1}{2}\r)}\frac{b^2}{2!}
+O(b^3) \r).
\ee
For $d=2$ an exact expression is given in Appendix \ref{App-d2}. It is useful
to note the near-boundary behavior of the scale factor obtained in
(\ref{warp-as}):
\be\label{nrbdy}
e^{2A(\mu)} \asympt_{\mu \to \pm \mu_0} \; \frac{1}{\rm{sin}^2(\mu_0 \mp \mu)}
[1 + O\l((\mu_0 \mp \mu)^{2d}\r)]
\ee
Thus the effect of the running dilaton on
the wall profile is a $b$-dependent change in the boundary limit $\mu_0$
together with an order $(\mu \pm \mu_0)^{2d}$ effect on the near
boundary shape.

\begin{figure}[!ht]
\begin{center}
\input{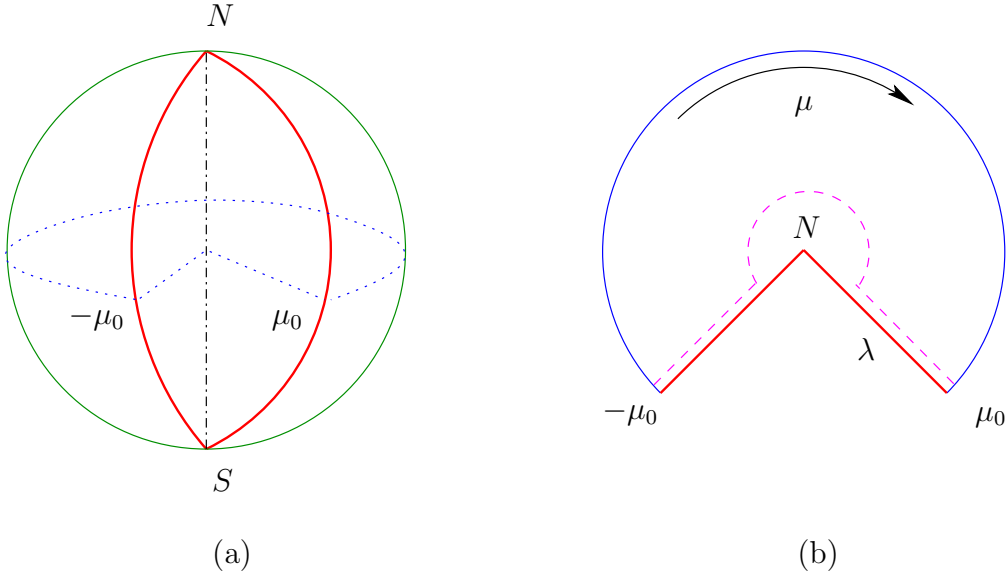}
\caption{\small (a) Conformal picture of a constant time slice of the
Janusian geometry. The boundary is indicated by the bold arcs. (b) Top view
of the same picture. The coordinate $\lambda$ ranges from $0$ at the equator
to $\pi/2$ at the north pole. The dashed line indicates the ``contour''  used
to evaluate $E_{WN}$ in Sec. 4. }
\label{Jan-pic}
\end{center}
\end{figure}

In (\ref{globmet}), which is the same as (20) of \cite{Gutp}, we have extracted
the conformal factor $e^{2A(\mu)}/\rm{cos}^2\lam$, so that the line element
in square brackets can be viewed, at least heuristically, as a conformal
compactification. As discussed in \cite{Gutp} this conformal metric is similar
to the Einstein static universe, and would agree with the well known conformal
compactification of $AdS_{d+1}$, in the limit $b \rightarrow 0$ when $A(\mu)
\rightarrow -\rm{\ln}(\rm{cos}\mu)$ and $\mu_0 \rightarrow \pi/2.$ In this
limit, the spatial metric (i.e. fixed $t$) is a hemi-sphere of $S_{d}$. For
$b>0$ and $\mu_0 > \pi/2$, we also have a half-sphere but with angular excess
as depicted in Fig.~\ref{Jan-pic}a.  The boundary of the conformal metric then
has two parts, hemi-spheres of $S_{d-1}$ at $\mu = \pm \mu_0$ which are joined
at the pole(s) where $\cos\lam=0$.

The angular coordinates $\mu,~\lam$ are singular at the poles. However, one may
choose regular coordinates there by embedding $S_{d}$ in ${\cal R}_{d+1}$ with
cartesian coordinates. For simplicity, we discuss the case $d=2$ in which we
take coordinates $z=\sin\lam,\; x=\cos\lam\cos\mu, \;
y=\cos\lam\sin\mu $. The induced metric on $S_{2}$, namely $d\bar{s}^2 =
dx^2+dy^2+dz^2~=~d\lam^2 +\cos^2\lam d\mu^2$, is then regular at the pole at
$x=y=0.$ Consider next the conformal factor $\Omega = e^{-A(\mu)} \cos\lam$.
It follows from (\ref{nrbdy}) that its near boundary behavior is
\bea
\Omega &\sim& \sin(\mu_0 \mp \mu) \cos\lam\\
   &\sim& x\sin\mu_0  \mp y \cos\mu_0
\eea
Thus $\nabla \Omega$, evaluated in the regular coordinates $x,y$ is
discontinuous as one continues from the boundary region $\mu=+\mu_0$, where
$y/x = +\tan(\mu_0)$, to the portion where $\mu=-\mu_0$ and where $y/x =
-\tan\mu_0$. This means that the factorization in (\ref{globmet}) does not
satisfy the strict definition of conformal compactification \cite{Penrose}). In
practice, it means that the conformal boundary has corners at the pole(s), a
geometric feature deduced by means of the regular cartesian
coordinates.\footnote{We are very grateful to Gary Gibbons for patient and
useful discussions of the geometry and its conformal compactification.} We will
treat the corner in the boundary integral that occurs in the Witten-Nester
stability analysis by deforming the contour around the corner as indicated in
Fig.~\ref{Jan-pic}b and taking the limit to the corner after the integration is
performed.

An alternative approach is to work with the
Fefferman-Graham coordinates, i.e. to look for a coordinate system where
the metric near the boundary takes the form
\be \label{FG}
ds^2 = {1 \over z^2}
\l[dz^2 + \l(g_{(0)ij} + z^2 g_{(2)ij} + \cdots
+ z^d \l(g_{(d)ij} + \log z h_{(d)ij}\r) + \cdots \r) dx^i dx^j \r].
\ee
Such coordinate system can always be reached \cite{FG}. In this expansion
$g_{(0)}$ is the boundary metric. All coefficients in (\ref{FG}) but $g_{(d)}$
are locally related to $g_{(0)}$. $g_{(d)}$ carries information about the
vacuum and correlation functions of the dual QFT, so this coordinate system is
well suited for holography \cite{deHaro}. Transforming the Janus solution to
this coordinate system appears laborious and seems to lead to singular
$g_{(d)ij}$. Since we will not address the AdS/CFT duality for this solution,
we will not present these results here and continue in the rest of the paper
with the coordinate system in (\ref{globmet}).

Domain walls with $AdS$ slicing can also be presented using a Poincar\'{e}
patch metric for the $AdS_d$ slices \cite{Gutp}. The scale factor $A(\mu)$ and
local aspects of the discussion above are not changed, but the global structure
is affected. In particular the metric is geodesically incomplete, and one needs
its global extension. For this reason we formulate our stability study using
the global version. The patch version may well be appropriate for the $AdS/CFT$
dual.

\section{Stability of flat domain walls}
\setcounter{equation}{0}
\label{Sec-Stability-flat}

We review in this section the stability argument \cite{Towns,SkenTowns} for
asymptotically $AdS$ ($AAdS$) flat domain walls with metric in the form
(\ref{domwall}) and accompanying scalar $\phi(r).$ The purpose of the argument
is to show that the energy of deformed solutions of the equations of motion
which approach the domain wall at large distance is higher than the energy of
the wall itself. We use the following notation for the background fields and
deviations:
\bea \label{deform}
ds^2 &=& [\bar{g}_{\mu\nu} + h_{\mu\nu}] dx^\mu dx^{\nu} \\
\phi &=& \bar{\phi} + \varphi.
\eea
The fluctuations $h_{\mu\nu},~\varphi$ are treated in full nonlinear fashion in
the interior of the spacetime, but they vanish on the boundary. We will not
state definite conditions on the boundary asymptotics in this section, but we
will be quite specific when we discuss the extension to $AdS_d$-sliced domain
walls in Sec.~\ref{Sec-AdS}.

The spinor formalism of Witten and Nester provides a generalized
``energy'' $E_{WN}$ with the following properties:
\begin{itemize}
\item It computes a linear combination of the conserved Killing charges
of the isometry group of the background, specifically the subalgebra contained
in the SUSY anti-commutator $\{Q,\bar{Q}\}$. Thus we expect to find spatial
translations in $d$ dimensions for flat walls and the charges of the algebra
$SO(d-1,2)$ for the $AdS_d$-sliced walls we treat in Sec.~\ref{Sec-AdS}.
\item
The charges vanish for the background solution under study,
i.e. when $h_{\mu\nu}$ and $\varphi$ vanish.
\item
The energy, defined with respect to the Killing vector
$\frac{\partial}{\partial t}$ of the background, is positive for
all fluctuations which are suitably damped on the boundary.
\end{itemize}

\subsection{The Witten-Nester energy and positivity}
To define $E_{WN}$, we consider a $d$-dimensional spacelike
surface $\Sigma$, which can be thought of as the initial value
surface for the Cauchy problem of the deformed domain wall spacetime.
Denote its boundary by $\partial \Sigma$. Then $E_{WN}$ is
defined by the boundary integral
\be \label{nestE}
E_{WN} =  \int_{\partial\Sigma} *\hat E
\ee
of the Hodge dual of the Nester 2-form
$\hat E = \frac{1}{2} \hat E_{\mu\nu} dx^\mu dx^\nu$, defined by
\be \label{3.nester}
\hat E^{\mu\nu} = \bar\eps_1 \, \Gamma^{\mu\nu\rho} \hat\nabla_\rho \eps_2
- \overline{\hat\nabla_\rho\eps_2} \, \Gamma^{\mu\nu\rho} \eps_1.
\ee
The covariant derivative is
\be \label{3.cov}
\hat\nabla_\mu = \nabla_\mu + W(\phi) \Gamma_\mu,
\ee
where $ W(\phi) $ is any function of $\phi$ which satisfies
\be \label{cond}
W(\phi) \asympt_{\phi \to \bar{\phi}} \; {1 \over 2 L } +  O\l( \varphi^2 \r).
\ee
The value of the integral over  $\partial
\Sigma$ depends only on the behavior of spinors $\eps_1(x),~\eps_2(x)$
near the boundary.  Thus, at this stage, the spinors can be arbitrary
in the interior, but must approach a background Killing spinor on
the boundary. In general one must take independent spinors
$\eps_1(x),~\eps_2(x)$ in order that $E_{WN}$ contain the full
set of background charges.

As in other treatments of gravitational energy, the surface
integral form of $E_{WN}$ is linear in the  fluctuations
$h_{\mu\nu},~\varphi$ and thus not manifestly positive. To establish
positivity we will use Stokes' theorem to rewrite (\ref{nestE}) as
an integral over $\Sigma$ and then impose more specific conditions
on $W(\phi)$ and $\eps_i(x).$ Stokes' theorem gives
\be \label{intsig}
E_{WN} = \int_{\Sigma} d\Sigma_\mu \nabla_\nu \hat E^{\mu\nu}.
\ee
Note that this step requires that the background and deformed solutions are
non-singular. There would otherwise be an additional surface contribution from
the singularity or the horizon which shields it. There are known methods
\cite{GHHP} to extend the treatment to include horizons, but naked
singularities present substantial new problems which are beyond the scope of
the present work. One should note also that the effective current $J^{\mu} =
\nabla_\nu E^{\mu\nu}$ is identically conserved, so $E_{WN}$ defines a
conserved quantity provided that boundary asymptotics of the integrand is
suitably restricted.

We now take $\eps_1=\eps_2$ in (\ref{3.nester}) because we are interested in
demonstrating positivity. The integrand of (\ref{intsig}) may now be
manipulated as in \cite{Towns} using the equations of motion (\ref{geneom}) and
regrouping of terms to obtain
\bea \label{hotstuff}
\label{3.vol}
E_{WN} &=& \int_\Sigma d\Sigma_\mu \left[ 2 \overline{\delta
\psi}_\nu \Gamma^{\mu\nu\rho} \delta \psi_\rho
- \frac{\kappa^2}{2} \overline{\delta \chi} \Gamma^\mu \delta \chi \right. \\
&& \left. + \bar{\eps} \Gamma^\mu \eps \left( -\kappa^2 V(\phi)+2(d-1)^2 \l(
\frac{1}{\kappa^2} W'^2 - \frac{d}{d-1} W^2 \r) \right) \right] \nonumber
\eea
where $W'=\pa_\f W$ and we have {\it defined}
\bea
\label{tino}
\delta \psi_\mu &=& \hat{\nabla}_\mu \eps  \\
\delta \chi &=& \l(\G^\mu \nabla_\mu \phi - {2 (d-1) \over \k^2} W'\r)\eps.
\label{dil}
\eea

The condition that $E_{WN}$ vanish in the undeformed background is satisfied
if we impose the following conditions:
\begin{itemize}
\item The last term in (\ref{hotstuff}) is cancelled, both with
and without fluctuations, if we require that
\be \label{VWrel}
V(\phi) =\frac{2(d-1)^2}{\kappa^2} \l( \frac{1}{\kappa^2} W'^2 - \frac{d}{d-1} W^2 \r).
\ee
\item We further require that (with no fluctuations)
\bea \label{bkgdkil}
\l(\bar{\nabla}_\mu +\bar\Gamma_\mu W(\bar{\phi})\r)\eps &=&0\\
\l(\bar{\Gamma}^\mu \bar{\nabla}_\mu \bar{\phi} -{2 (d-1) \over \k^2}
W'(\bar{\phi})\r)
\eps &= &0.
\eea
An ``overbar'' on  any quantity indicates that it is to be
evaluated in the background geometry.
\end{itemize}

The integrability conditions of the equations (\ref{bkgdkil}), expressed in the
coordinates of (\ref{domwall}), give the first order flow equations
(\ref{flowA}, \ref{flowphi}) \cite{SkenTowns}.
The spinor solutions of (\ref{bkgdkil}) are the
background Killing spinors
\bea \label{kilspin}
\eps &=& e^{\frac{A(r)}{2}} \eps_0  \nonumber
\\
\Gamma^{\hat{r}} \eps_0 &=&\eps_0,
\eea
where $\eps_0$ is a constant spinor which is chiral with respect to the radial
component of $\Gamma^a.$

The superpotential $W(\phi)$ must satisfy
(\ref{VWrel}) and the boundary condition (\ref{cond}).
This guarantees that the
scalar profile $\bar{\phi}(r)$ and scale factor $A(r)$ obtained from
(\ref{flowphi},~\ref{flowA}) are also solutions of the field equations
(\ref{dweom}).

Positivity with fluctuations is now a relatively simple matter,
since there are only two terms left in (\ref{hotstuff}) and the
second one is manifestly positive.
We still have the freedom to modify the definition of the spinor
$\eps(x)$ for deformed solutions, and we impose the
Witten condition
\be \label{wit}
\Gamma^{k} \nabla_{k} \eps(x) =0
\ee
where the time coordinate is omitted in the sum over $k$. One must choose a
solution which approaches an arbitrary background Killing spinor on the
boundary. We do not discuss the existence of Witten spinors here. In a frame
where $E^{\hat t}$ is orthogonal to the surface $E_{WN}$ reduces to the
positive semi-definite form
\be
E_{WN} = \int_\Sigma d^{d{-}1}\! x\, e \left[ 2
\l(\hat{\nabla}^{k}\eps\r)^\dagger \hat{\nabla}_{k}\eps + \frac{\kappa^2}{2}
\delta \chi^\dagger \delta \chi \right].
\ee
This energy functional vanishes if and only if
\be
\hat{\nabla}_{k}\eps=0, \qquad \delta \chi=0.
\ee
The set of solutions to these equations is given by the space of
solutions of the first order equations (\ref{flowA}, \ref{flowphi}). If we
impose that the solution should satisfy the boundary condition
set by the undeformed solution, then $E_{WN}$ vanishes only for the undeformed
background. The existence of other solution that have
zero energy but different boundary conditions may be considered
as an indication of marginal stability, but it is unclear whether
we should allow such configurations. We leave this issue open.

We can now state a sufficient condition for the stability of a domain
wall of the form (\ref{domwall}) which is an asymptotically $AdS$ solution
of (\ref{dweom}) and involves a scalar whose mass satisfies
$m^2_{BF} \equiv -\frac{d^2}{4} \le m^2 \le 0.$ The
scalar profile then satisfies  $\bar{\phi} ' =0$ on the boundary.
If there is a superpotential $W(\phi)$, satisfying (\ref{VWrel}),
such that the domain wall is a solution of (\ref{flowphi},~\ref{flowA}),
then it is stable. It then follows from (\ref{flowphi}) that $W'$ vanishes
on
the boundary. For analytic $W(\phi)$ this is equivalent to (\ref{cond}).
It is not guaranteed that the required adapted superpotential exist.
As discussed at the end of Subsection 2.1, if
$\bar{\phi}(r)$ is monotonic, then $W(\phi)$ is defined implicitly.
If it doesn't exist, then one may suspect instability, but
instability does not follow from this framework.

The roots of the argument above lie in supergravity, as the matrix structure of
the Nester 2-form (\ref{nestE}) and the form of (\ref{tino}, \ref{dil}) clearly
show. But the argument can be applied to any model of gravity and scalar
fields, in any space-time dimension, provided that the required adapted
superpotential exists.

\subsection{$E_{WN}$ and conserved charges}

Our next goal is to obtain a concrete formula for the boundary integral form of
$E_{WN}$ and to show that it indeed gives a combination of the translation
Killing charges of flat domain walls. Because we work at the boundary,
linearized expressions for the connection and frames of the deformed metric are
appropriate. Linearization in the scalar fluctuation is valid for single-scalar
models where the scalar mass satisfies $m^2 > m^2_{BF}$. However, terms of
order $\varphi^2$ can contribute to $E_{WN}$ when the scalar mass saturates the
BF bound \cite{brfr} and in other situations. An example was recently discussed
in \cite{Hertog}.

Let $\bar{E}^a_\mu$ denote a vielbein of the
background metric in (\ref{deform}). The linearized spin
connection is then given by
\be \label{conn}
\delta\omega_{\mu a b} = \frac{1}{2} \l[ \bar{E}_a^\nu \nabla_b h_{\mu\nu}  -
\bar{E}_b^\nu \nabla_a h_{\mu\nu} \r]
\ee
where $\nabla$ is a background covariant derivative. It is most
convenient to use background Killing spinors to compute $E_{WN}$.
We insert (\ref{conn})
in (\ref{nestE}) and obtain, using (\ref{bkgdkil}) and some Dirac algebra,
\bea \label{truth}
E_{WN} &=& -\frac{1}{8} \int_{\partial\Sigma} \l[ \bar\eps_1 \Gamma_\mu
\eps_2 \l(\nabla^\rho h_{\rho\nu} - \nabla_\nu \l( g^{\rho\sigma}
h_{\rho\sigma} \r)\r) - \bar\eps_1 \Gamma_\nu \eps_2
\l(\nabla^\rho
h_{\rho\mu} - \nabla_\mu \l( g^{\rho\sigma} h_{\rho\sigma} \r)\r)
 \r.
\nonumber\\
&& \qquad\qquad + \l. \bar\eps_1 \Gamma^\rho \eps_2 \l(\nabla_\nu
h_{\rho\mu} - \nabla_\mu h_{\rho\nu} \r) \r] d\Sigma^{\mu\nu}
\nonumber\\
&& + \frac{1}{4} \int_{\partial\Sigma} \l[ \bar\eps_1 \l( g^{\mu\sigma}
\Gamma^{\nu\rho} + g^{\nu\sigma} \Gamma^{\rho\mu} + g^{\rho\sigma}
\Gamma^{\mu\nu} \r) h_{\rho\sigma} W(\bar{\phi}) \eps_2 \r] d\Sigma_{\mu\nu}
\nonumber\\
&& + \frac{d-1}{2} \int_{\partial\Sigma} \bar\eps_1 \Gamma^{\mu\nu}
W'(\bar{\phi}) \eps_2 \varphi \, d\Sigma_{\mu\nu} + h.c.
\eea
All quantities in this equation, except $h_{\mu\nu}$ and $\varphi$
refer to the background.
Our computation used only the general background-fluctuation
split in (\ref{deform}). It is thus valid both for flat domain
walls and for other situations in which the Witten-Nester
approach to stability has been applied. For example, it is
applicable to asymptotically flat metrics in which $W(\bar{\phi})$
and $W'(\bar{\phi})$ vanish. In this case it is quite
straightforward to show that (\ref{truth}) yields the same
expressions for energy and momentum given in (70) of \cite{Witten}.

Let us discuss the formula (\ref{truth}) in more detail for flat domain
walls. First we find from (\ref{kilspin}) that the bilinears
$\bar{\eps}_1 \Gamma^\mu \eps_2$ do span the expected set of
translation Killing vectors\footnote{The case d=2 is
exceptional. Due to chirality, the Killing spinors have
effectively only one component, so $\bar{\eps}_1 \Gamma^\mu
\eps_2$ has vanishing spatial component and gives only the time
translation or energy Killing vector.}
However, the role of tensor bilinears
$\bar{\eps}_1 \Gamma^{\mu\nu} \eps_2$ is far from clear. To
discuss them, we distinguish between components
$\bar{\eps}_1 \Gamma^{ri} \eps_2$ with one radial index, and
components $\bar{\eps}_1 \Gamma^{ij} \eps_2$ with both indices
along the domain wall. The latter vanish due to the chirality
properties of the Killing spinors (\ref{kilspin}), while the
former are proportional to translation Killing vectors. Thus $E_{WN}$
indeed produces a combination of the translation Killing charges
of the deformed domain wall! We note further that  $W'(\bar{\phi})$
vanishes, so that the last term in (\ref{truth}) is absent for
flat domain walls.

As a final check let us note that the boundary volume element has components
$d\Sigma_{tr}$ where $t$ is the time coordinate of (\ref{domwall}).  We now use
radial coordinate $r$ for which $h(r)=0$ in (\ref{domwall}). In that case,
$A(r) \sim r/L$ at the boundary. It is also known that normalizable metric
fluctuations vanish at the rate $h_{\mu\nu}\sim \exp(-dr/L)$. Putting things
together we see that the terms in the first three lines of (\ref{truth}) are
generically finite on the boundary.

We conclude this section with an illustration of one of the subtleties of the
argument, namely that the existence of an adapted superpotential satisfying
(\ref{VWrel}) is not
sufficient to guarantee stability. In addition one needs (\ref{cond}) which
implies that the AdS critical point of the potential $V$ is also a critical
point of $W$. To illustrate this issue we consider the following
superpotential:
\be
W = w_0 + w_1 \phi + \frac{d}{2(d-1)} \kappa^2 w_0 \phi^2 + w_3 \phi^3
\ee
The corresponding potential from (\ref{VWrel}) is
\be
\kappa^2 V(\phi) = 2 (d-1)^2
\left({w_1^2 \over \kappa^2} - {d \over (d-1)} w_0^2\right)
-2 (d-1) \left(d w_1^2 - {6 \over \k^2} w_1 w_3 (d-1)\right) \phi^2
+  O(\phi^3).
\ee
This potential has a critical point at $\phi=0$ which is $AdS$ provided
\be
w_1^2 < {d \over (d-1)} \kappa^2 w_0^2.
\ee
This critical point however is not a critical point of $W$. If the product
$w_1w_3$ is sufficiently large, the mass of the scalar lies below $m^2_{BF}$
and the perturbative argument \cite{brfr} for instability applies. We may apply
the Witten-Nester argument to investigate stability of the $AdS$ solution of
the theory (\ref{toymod}) with potential above. The argument does not apply if
one uses the covariant derivative (\ref{3.cov}) with $W$ above because AdS
spacetime is not a solution\footnote{A preliminary study indicates that the
flow equations can be integrated, but give a pathological geometry.}  of the
flow equations (\ref{flowphi},~\ref{flowA}). Nor can there be any other
superpotential, satisfying both (\ref{VWrel}) and (\ref{cond}) because it is
known \cite{Towns} that this implies that $m^2 \ge m^2_{BF}.$ Thus the
perturbative and non-perturbative analysis are compatible. This example
illustrates the importance of the condition (\ref{cond}) for stability.

\section{Stability of $AdS_d$ domain walls}
\setcounter{equation}{0} \label{Sec-AdS}

In this section we extend the argument of Sec. 3 to cover $AdS_d$-sliced domain
walls. The springboard for our approach was the study of $AdS_4$-sliced walls
in genuine $D=5,~~N=2$ supergravity in \cite{Lust}. The natural spinors in this
theory are a symplectic-Majorana doublet, and the superpotential appears as the
$su(2)$-valued matrix ${\bf W}(\phi) = W_a(\phi) \tau^a$, where the $\tau^a$
are the three Pauli matrices.   
In genuine $D=5,~N=2$ supergravity, the matrix superpotential 
is determined by the gaugings of R-symmetry and isometries of the
internal geometry \cite{Ceresole1,gz,Ceresole2}. The internal space is the product 
of a very special manifold (for
scalars in vector and tensor multiplets) and  a
quaternionic manifold (for scalars in hypermultiplets).
The superpotential is given by the product of the embedding coordinates
$h^I$ of the very special manifold and a triplet of Killing prepotentials
$P_{Iij}$ depending on the scalars of the hypermultiplets.
In the absence of hypermultiplets, a matrix superpotential is still 
possible\footnote{We thank Antoine Van Proeyen for correspondence on this issue.} 
and it is determined in terms of Fayet-Iliopoulos 
constants and the $h^I$.

None of this technical detail need concern us
in fake supergravity, which works in any dimension and
with any number of real scalars. We simply double the spinors
used in Sec. 3, taking $\e^{\a},~~\a =1,2$ as a pair of Dirac
spinors in dimension $d+1$. The matrix ${\bf W}(\phi)$ acts on
the index $\a$, but we can usually suppress it in explicit formulas.
Many previous formulae remain valid when understood as extensions
to the doubled spin space, with the replacement
$W(\phi) \rightarrow {\bf W}(\phi)$. Note that quadratic quantities
such as $ {\bf W}^2$ and $\{{\bf W},{\bf W'}\}$ are proportional
to the unit matrix. When they appear in our equations below they
should be interpreted as scalar-valued.

The energy of any perturbation of an $AdS_d$-sliced wall is contained in the
Nester 2-form (\ref{3.nester}) with an $su(2)$ extension of the covariant
derivative (\ref{3.cov}). All formal manipulations which lead to the volume
form (\ref{3.vol}) of the energy also have obvious $su(2)$ extensions. With an
$su(2)$-extended Witten spinor (\ref{wit}), the energy becomes manifestly
non-negative. The non-trivial task now is to establish the consistency of the
formalism by showing that there are fake Killing spinors so that the energy
vanishes for domain wall backgrounds of the form (\ref{adswall}). We use the
frames and spin connections given in Appendix A.

\subsection{Killing spinor consistency conditions and the new
flow equations.}
The $su(2)$ extension of the argument of Sec. 3 requires that the
fake Killing spinors satisfy the following conditions\footnote{
As in Sec. 3, $\W'$ and $\W''$ denote derivatives with respect to $\phi$.}:
\bea
\label{fake1}
\l[ \nabla_i^{AdS_d} + \Gamma_i \l(\frac{1}{2} A' e^{-h}
\Gamma^{\hat{r}} +{\bf W} \r) \r] \eps &=& 0
\\
\label{fake2}
\l[ \partial_r + \Gamma^{\hat{r}} e^{h} {\bf W} \r] \eps &=& 0
\\
\label{fake3}
\l[ \Gamma^{\hat{r}} e^{-h} \phi' - \frac{2(d-1)}{\kappa^2} {\bf W'} \r] \eps &=& 0.
\eea
In addition ${\bf W}(\phi)$ must be related to the potential
$V(\phi)$ by
\be \label{VWads}
\kappa^2 V(\phi) = 2(d-1)^2 \l( \frac{1}{\kappa^2} {\bf
W'}^2 - \frac{d}{d-1} {\bf W}^2 \r).
\ee
In (\ref{fake1}), the covariant derivative contains the
connection of an $AdS_d$ metric with scale $L_d$.

We now extract the integrability/consistency conditions for (\ref{fake1} --
\ref{fake3}) and show that they imply that the background metric and scalar
satisfy the original Euler-Lagrange equations  (\ref{adseom}). We also obtain a
constraint on ${\bf W(\phi)}$.

Consider first the fake dilatino condition (\ref{fake3}) which
can be rewritten as the chirality condition
\be \label{chir}
\Gamma^{\hat{r}} \eps
=  \frac{2(d-1)}{\kappa^2} e^h  \frac{{\bf W'}}{\phi'} \eps
\ee
on fake Killing spinors. The square of this gives the scalar condition
\be \label{phicon}
{\phi'}^2 - \l(\frac{2(d-1)}{\kappa^2}\r)^2 e^{2h}
{\bf W'}^2 = 0,
\ee
which shows that the matrix on the right side of (\ref{chir}) has
eigenvalues
$\pm1$, as required for the consistency of (\ref{chir}).

The integrability condition for (\ref{fake1}) is
\be
\label{intcon1}
\frac{1}{L_d^2} + {A'}^2 e^{2A-2h} - 4e^{2A} {\bf W^2} = 0,
\ee
while the compatibility of (\ref{fake1}) and (\ref{fake3})
requires (after use of (\ref{chir}))
\be
\label{intcon2}
A' \phi' + \frac{2(d-1)}{\kappa^2} e^{2h} \{{\bf W}, {\bf W'} \} = 0,
\ee
The mutual integrability condition for (\ref{fake1}, \ref{fake2}) directly
gives the $A'' - A'h'$ field equation of (\ref{adseom}) after (\ref{phicon})
and (\ref{intcon2}) are used. The remaining compatibility condition between
(\ref{fake2}, \ref{fake3}) will be discussed below. It is an important
constraint on ${\bf W}(\phi)$.

We can now easily recover the other equations of motion in
(\ref{adseom}). First we combine (\ref{phicon}) and
(\ref{intcon1}) and use (\ref{VWads}) to obtain the $A'^2$
equation from (\ref{adseom}). Next we take the $r$-derivative of
(\ref{phicon}) and find
\be \label{phi2nd}
\phi'' -h' \phi' =  \frac{2(d-1)^2}{\kappa^4} e^{2h} \{ \W', \W''\}.
\ee
The sum of this plus $d$ times (\ref{intcon2}) yields exactly the
scalar equation in (\ref{adseom}).
Our formalism is thus consistent with
the field equations of $AdS_d$-sliced domain walls!

The next step is to extract from the information above
a small set of equations which determine
$\phi(r),~A(r)$.
The first equation is just the square root of (\ref{phicon}) with
sign chosen to make $\phi(r)$ monotonically increasing 
\be \label{phiflow}
\phi'(r) = \frac{2(d-1)}{\kappa^2} e^{h}
\sqrt{\W'^2}\,.
\ee
The second equation is a purely algebraic equation for $A(r)$,
obtained by equating the expressions for $A'^2$ obtained from (\ref{intcon1})
and from (\ref{intcon2},~\ref{phiflow}):
\be \label{Aalg}
\frac{e^{-2A}}{L_d^2} ~=~\frac{4{\bf W}^2{\bf W'}^2 - \{{\bf
W},{\bf W'}\}^2}{{\bf W'}^2}.
\ee
The right side is non-negative by the Schwarz inequality.

We now show that (\ref{phiflow},~\ref{Aalg}) are equivalent to the first order
set  (\ref{dfgkflow}) provided that ${\bf W}$ satisfies (\ref{VWads}) and a
further condition given below. This then guarantees that the new system gives a
solution of the original field equations (\ref{adseom}). In making the
comparison with (\ref{dfgkflow}), we interpret $W =\sqrt{{\bf W}^2}$ and $W' =
\frac{d}{d\phi} W$. First we must require that the relations between ${\bf W}$
and the potential $V(\phi)$ in (\ref{adseom}) and (\ref{VWads}) are equivalent.
Thus we identify
\be \label{gamsq}
\gamma^2 = \frac{\{{\bf W},{\bf W'}\}^2}{4 {\bf W}^2{\bf W'}^2}.
\ee
The algebraic equation (\ref{Aalg}) then implies (\ref{gamdef}).
\footnote{For the Janus solution discussed further in Sec. 4.3. the factor $\gamma$ 
appearing in (\ref{gamdef}) vanishes at $r=0$ and has to be therefore extended as an 
odd function to negative $r$. This amounts to setting
$\gamma = -\frac{\{{\bf W},{\bf W'}\}}{2 \sqrt{\W^2\W'^2}}$.
} 
This also shows that
(\ref{phiflow}) is equivalent to the $\phi'$ equation in  (\ref{dfgkflow}).

It is also easy to obtain the $A'$ equation in
(\ref{dfgkflow}). Substitute (\ref{phiflow}) into (\ref{intcon2})
which gives
\be \label{newA}
A' = -\frac{\l\{\W, \W'\r\}}{\sqrt{{\bf W'}^2}}e^h.
\ee
We then use (\ref{gamsq}) to recover the form in (\ref{dfgkflow}).
However, there is a subtlety here. Namely (\ref{newA}) is compatible
with the expression for $A'$ obtained from the logarithmic
derivative of (\ref{Aalg}) combined with (\ref{phiflow}) only if
${\bf W}(\phi)$ satisfies the constraint
\be\label{Wcon1}
\frac{\Tr \W \W' \, \Tr \W' \W'' - \Tr {\W'}^2  \, \Tr \W \W''}{
\Tr \W^2  \, \Tr {\W'}^2 - \l(\Tr \W \W' \r)^2} =
\frac{\kappa^2}{d-1}.
\ee

The compatibility condition between (\ref{fake2}) and (\ref{fake3}) provides a
simple direct constraint on the superpotential ${\bf W}(\phi)$ which supersedes
(\ref{Wcon1}). After use of (\ref{phi2nd}) and (\ref{chir}), we find that ${\bf
W}$ must obey the following consistency condition
\be \label{Wcon2}
\l[{\bf  W'},\frac{d-1}{\kappa^2} {\bf  W''} + {\bf  W} \r] =0.
\ee
This condition, which must hold for any potential, is a necessary
condition for the existence of fake Killing spinors and will be
important in their construction below.

Since the Cartan subalgebra of $su(2)$ is one dimensional,
\be\label{plane}
\W'' = \alpha(\phi)\, \W' - \frac{\kappa^2}{d-1} \W,
\ee
where $\alpha(\phi)$ is a real function of the scalar
field. One can see that (\ref{Wcon1}) is trivially satisfied if
(\ref{plane}) is inserted.
By taking the anti-commutator of both
sides of (\ref{plane}) with $\W'$, one finds that
\be\label{def_alpha}
\alpha(\phi) = \frac{\kappa^2}{2(d-1)\, {\W'}^2} \l[(d+1)\l\{\W,\W'\r\} + \frac{\kappa^2}{2(d-1)}
\frac{\partial V}{\partial \phi} \r].
\ee

The equation (\ref{plane}) implies that the matrix ${\bf W''}$ lies in the
vector space spanned by matrices ${\bf W}$ and ${\bf W'}$.
Taking further derivatives one can see that actually all derivatives
lie in the same two dimensional vector space. Thus, assuming
analyticity, the superpotential
${\bf W}(\phi)$ remains in a fixed subspace for all values of
$\phi$. This allows us to make the convenient gauge choice
\be
\W = \l(\begin{array}{cc} 0 & \bar\omega \\
                      \omega & 0
    \end{array}\r).
\ee
In this gauge the consistency condition (\ref{Wcon2}) reduces
to
\be\label{omegacon}
\frac{\bar\omega' \omega'' - \omega' \bar\omega''}{\bar\omega \omega' - \omega \bar\omega'}
= \frac{\kappa^2}{d-1}.
\ee

It is quite remarkable that we have replaced the system
(\ref{dfgkflow}) by the simpler set (\ref{phiflow},~\ref{Aalg}) in which
only one integration is required given the superpotential ${\bf
W}(\phi)$. However, the conditions (\ref{VWads},~\ref{plane},~\ref{def_alpha})
which determine  ${\bf W}(\phi)$ from $V(\phi)$ are not
necessarily easy to solve, as we discuss below. It
appears possible to shift the strategy as follows. First obtain a
superpotential which satisfies (\ref{plane}) and use
(\ref{def_alpha}) to define a potential. The $AdS_d$-sliced
domain wall then obtained from  (\ref{phiflow},~\ref{Aalg}) will
be stable.

We may summarize the results above as follows. If the matrix superpotential
${\bf W}(\phi)$ satisfies (\ref{Wcon2}) and (\ref{VWads}), then any solution of
(\ref{phiflow},~\ref{Aalg}) satisfies the field equations (\ref{adseom}) for
$AdS_d$-sliced domain walls. The Killing spinor equations (\ref{fake1} --
\ref{fake3}) are then mutually consistent and we should be able to find the
Killing spinors.

\subsection{Explicit Killing spinors.}
Let $\eps_K$ denote a conventional
Killing spinor of $AdS_d$ which satisfies\footnote{In this equation 
$\Gamma_i = \bar e_{i \hat a} \Gamma^{\hat a}$ is an $AdS_d$ gamma matrix. } 
\be
\l[ \nabla_i^{AdS_d} + \frac{1}{2L_d}\Gamma_i \r]\eps_K =0.
\ee
For $d=4$ there are 8 independent $\eps_K$. For each independent $\eps_K$,
there is an $su(2)$ fake Killing spinor of the form,
\be\label{fake_Killing_sol}
\eps = e^{\frac{1}{2} A}
     \l(\begin{array}{cc} \frac{\bar\omega'}{\omega'} & 0 \\
                      0 & \frac{\omega'}{\bar\omega'}
     \end{array}\r)^{\frac{1}{4}}
  \l(\begin{array}{c} (i + \Gamma^{\hat r}) \eps_K \\
                      -(1 + i \Gamma^{\hat r}) \eps_K
    \end{array}\r).
\ee
One can check directly that the defining conditions (\ref{fake1}) --
(\ref{fake3}) are satisfied. For this purpose one needs the following formulas:
\bea \label{nofake1}
\frac{2A'}{\phi'} &=&
-\frac{\kappa^2}{d-1} \l(\frac{\omega}{\omega'}+\frac{\bar\omega}{\bar\omega'}\r)
\\\label{nofake2}
\partial_\phi\l(\log\frac{\omega'}{\bar\omega'}\r) &=& -\frac{\kappa^2}{d-1}
\l(\frac{\omega}{\omega'}-\frac{\bar\omega}{\bar\omega'}\r)
\\\label{nofake3}
0 &=&-i \frac{e^{-A}}{L_d} + A'e^{-h} - 2 \bar{\omega}
\sqrt{\frac{\omega'}{\bar{\omega}'}}
\eea
which follow easily from (\ref{phicon})--(\ref{intcon2}) and (\ref{omegacon}).
Note that the prime on $\omega$ and $\bar\omega$ means a derivative with
respect to $\phi$, whereas the prime on $A$ or $\phi$ means a derivative with
respect to $r$. The fake Killing spinor bilinears $\bar{\eps}_1 \gamma^i \eps_2$ (with
$\eps_1 \ne \eps_2$) span the set of Killing vectors of the $AdS_d$ isometry
group $SO(d-1,2)$, as they should.

\subsection{${\bf W}(\phi)$ for the Janus solution.}

In this subsection we analyze the conditions which determine ${\bf
W}(\phi)$ in more detail and show that there is a solution which generates the
solution of \cite{Gutp} and thus establishes its stability.

Inserting the ansatz
\be\label{w-ansatz}
\omega(\phi) =  w(\phi) \, e^{i \theta(\phi)}
\ee
into (\ref{omegacon}) and (\ref{VWads}) one finds
\bea\label{w_theta_system}
 w'^2+w^2 \theta'^2  - \frac{d\,\kappa^2}{d-1} w^2 &=& \frac{\kappa^4}{2(d-1)^2} V(\phi),
\nonumber\\
\frac{2 w'^2}{w^2} + \frac{\theta''}{\theta'}\frac{w'}{w} -\frac{w''}{w}+\theta'^2 &=&
\frac{\kappa^2}{d-1} \qquad \mbox{or} \quad \theta'=0.
\eea
Eliminating $\theta$ from the system of equations we find
\be\label{Xeq}
X X'' - \frac{d+1}{2d} X'^2 + \frac{d+2}{2d} \kappa^2 V' X' - \kappa^2 \l(V'' + \frac{2\kappa^2}{d-1} V \r) X
- 2\kappa^2 X^2 = \frac{\kappa^4}{2d} V'^2,
\ee
where we have introduced
\be \label{Xdef}
X(\phi) = 2d(d-1) w^2 + \kappa^2 V.
\ee
For a {\it constant} potential $V=V_0$ this is an autonomous differential
equation which can be solved by standard methods.\footnote{For an exponential
potential, we can write $X=V Y$ and obtain again an autonomous equation for
$Y$.}
One takes as a new independent variable $X$ and new dependent variable $u=X'$. Then using
$\frac{d}{d\phi}=u\frac{d}{dX}$ we find a first order linear ODE for $u^2$,
\be\label{XODE}
\frac{X}{2} \frac{d}{dX} \l(u^2\r) - \frac{d+1}{2d}u^2 - \frac{2\kappa^4}{d-1} V_0 X - 2\kappa^2 X^2 =0
\ee
Solving this equation and passing back to the original variables we find $X(\phi)$ defined implicitly by
\be \label{Xphi}
\sqrt{\frac{d-1}{4d\, \kappa^2}} \int_{0}^{X}\! \frac{dx}{\sqrt{x^2-\kappa^2
V_0 x - \beta x^{\frac{d+1}{d}}}} = \phi_{\infty} - \phi,
\ee
where $\phi_\infty$ is the boundary value of the field at $r=+\infty$.
We have fixed the shift invariance of (\ref{XODE}) by requiring that $X(\phi=0)=x_{min}$,
where $x_{min}$ is the smallest positive root of the denominator in (\ref{Xphi}).
Equation (\ref{Xphi}) thus defines $X(\phi)$ for $\phi \ge 0$ only. 
It can be continued, however, as an even $C^\infty$ function to negative $\phi$.  

\begin{figure}[th]
\leavevmode
\begin{center}
\input{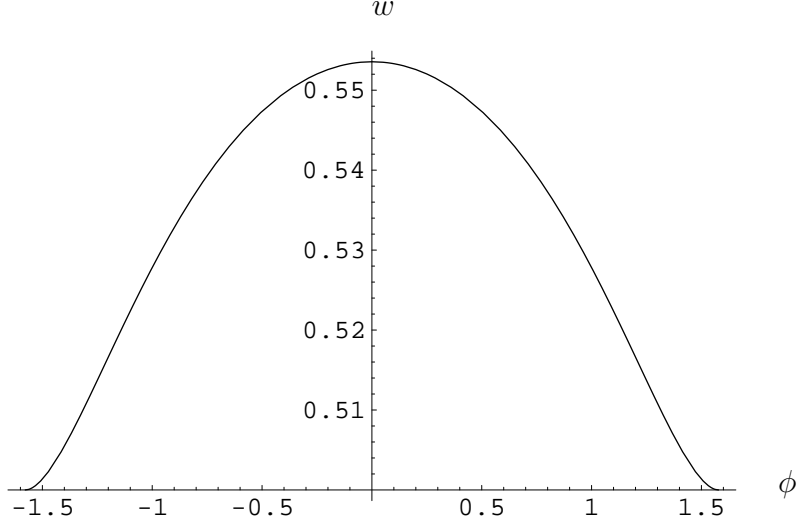}
\end{center}
\caption{\small Plot of the magnitude $w(\phi)$ for $d=4$, $L=1$ and $b=0.1$.}
\label{w_plot}
\end{figure}

\begin{figure}[th]
\leavevmode
\begin{center}
\input{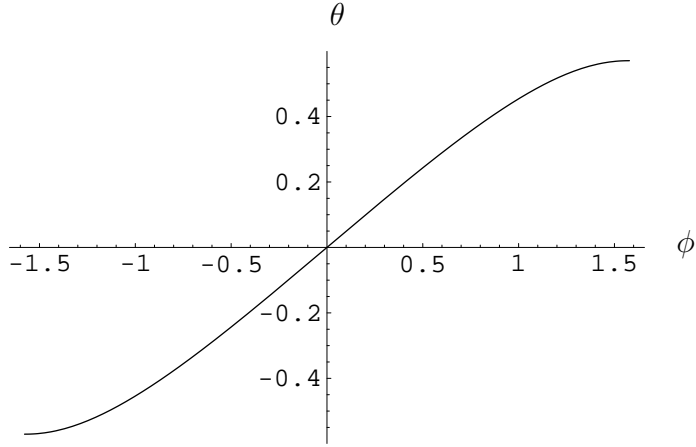}
\end{center}
\caption{\small Plot of the phase $\theta(\phi)$ for $d=4$ and $b=0.1$.}
\label{th_plot}
\end{figure}

As we shall show below the
integration constant $\beta$ is related to the parameter $b$ of the Janus
solution by equation (\ref{b-beta}). Once we have obtained the magnitude $w$ we
can find the phase $\theta$ simply by an integration
\be\label{thetaphi}
\theta_\infty -\theta = \frac{\sqrt{\beta}}{2} \int_{0}^{X}\! \frac{dx \;
x^{\frac{d+1}{2d}}}{(x-\kappa^2 V_0 )\sqrt{x^2-\kappa^2 V_0 x - \beta
x^{\frac{d+1}{d}}}}.
\ee
From (\ref{Xdef},~\ref{Xphi},~\ref{thetaphi}) one can find the behavior of the
superpotential as $\phi \rightarrow \phi_\infty$, namely
\bea \label{phib}
X &\simeq& \frac{\kappa^2 d^2}{2L^2} (\phi-\phi_\infty)^2,
\nonumber\\
w &\simeq& \frac{1}{2L} + \frac{\kappa^2 d}{4L (d-1)} (\phi-\phi_\infty)^2,
\nonumber\\
\theta &\simeq& \theta_\infty - \frac{\sqrt{\beta}\, d}{2d+1}
\l(\frac{2L^2}{d(d-1)}\r)^{\frac{3}{2}} \l(\frac{\kappa^2
d^2}{2L^2}\r)^{1+\frac{1}{2d}} |\phi-\phi_\infty|^{2+\frac{1}{d}},
\eea
Plots of the magnitude and the phase of the superpotential are shown in Figures
\ref{w_plot}, \ref{th_plot}.

Let us now demonstrate that the above fake superpotential does indeed generate
the Janus solution. From the definition (\ref{Xdef}) and the relation
(\ref{VWads}) we find easily
\bea
\W^2 &=& \frac{1}{2d(d-1)} X + \frac{1}{4L^2},
\\\label{W'Xrel}
\W'^2 &=& \frac{\kappa^2}{2(d-1)^2} X,
\\
\l\{ \W, \W' \r\} &=& \frac{X'}{2d(d-1)}.
\eea
The scale factor can then be calculated from (\ref{Aalg}) and (\ref{Xphi})
\be\label{Aalg-Jan}
e^{-2A} = L_{d}^2 \frac{2\beta}{d(d-1)} X^{\frac{1}{d}}.
\ee
To facilitate the comparison let us choose a coordinate in which the dilaton is
linear in the coordinate $r$. In particular we take $\phi(r) =\frac{r}{\kappa
L}$. Clearly this can be achieved for the Janus solution since the dilaton is a
monotonic function of the radial variable. Using $\phi' =\frac{1}{\kappa
L}$ we find using (\ref{phiflow}) and (\ref{W'Xrel})
\be\label{hX}
e^{-2h} = 2 L^2 X.
\ee
From (\ref{Aalg-Jan}) and (\ref{hX}) we see that $h_0 \equiv h-dA$ is a constant and is given by
\be
e^{2h_0} = \frac{1}{2 L^2}  \l( \frac{2\beta L_{d}^2}{d(d-1)} \r)^{d}.
\ee
Taking a logarithmic derivative of (\ref{Aalg-Jan}) we find
\bea\label{Aprime-Jan}
A' &=& -\frac{1}{2d} \frac{X'}{X} \phi'
\nonumber\\
&=& \frac{1}{L} \sqrt{ \frac{1}{d(d-1)} + e^{2dA + 2h_0} -
\l(\frac{L}{L_d}\r)^2 e^{2(d-1)A + 2h_0}}.
\eea
Comparing this first order ODE with the equation obeyed by the Janus solution
following from (\ref{adseom}) in the same linear dilaton coordinates,
we see that they are indeed the same provided we identify
\be
\label{b-beta}
\beta = \frac{d(d-1)}{2L_d^2} \l( \frac{2L^2}{b\, d (d-1)} \r)^{\frac{1}{d}}.
\ee
Note that the coordinate independent definition of $b$ is
\be
b = \frac{\kappa^2 L^2}{d(d-1)} {\phi'}^2 e^{2dA - 2h},
\ee
which is indeed a spacetime constant as follows from the equation of motion in
(\ref{adseom}).

Finally let us mention, that in addition to the Janus solution, there are
other simpler solutions to the equations  (\ref{w_theta_system}).
In particular there are two solutions with constant magnitude
\bea \label{lindil}
X &=& 0, \qquad \quad w^2 = \frac{1}{4L^2}, \qquad \qquad \theta(\phi) = const.
\nonumber\\
X &=& \frac{d}{2L^2}, \qquad w^2 =  \frac{1}{4L^2} \frac{d}{d-1},
\qquad \theta(\phi) = \pm \frac{\kappa}{\sqrt{d-1}} \, \phi.
\eea
The first solution is just the standard $AdS_{d+1}$ space, whereas
the second solution leads to an interesting linear dilaton
background discussed further in Section \ref{Section_LinearDilaton}.
The equation (\ref{Xeq}) also admits a cosh type solution
\bea
X &=& \frac{d(d-1)}{2L^2} \sinh^2 \l(\kappa\sqrt{\frac{d}{d-1}} \l(\phi-\phi_0\r) \r)
\nonumber\\
w^2 &=& \frac{1}{4L^2} \cosh^2 \l(\kappa\sqrt{\frac{d}{d-1}} \l(\phi-\phi_0\r) \r)
\nonumber\\
\theta &=& \theta_0.
\eea
However, (\ref{phiflow}) and (\ref{newA}) then generate the singular profiles
found for flat dilaton walls in Sec. 2.1. This case appears to be a degenerate 
limit of our equations,
since the right hand side of (\ref{Aalg}) vanishes, implying that $L_d \to \infty$.

\subsection{$E_{WN}$ for deformations of the Janus solution.}

We have demonstrated above the existence of an $su(2)$ superpotential ${\bf
W(\phi)}$ for which (\ref{phiflow},~ \ref{Aalg}) generate the $AdS_4$-sliced
domains wall of \cite{Gutp} and its $d$-dimensional generalizations. This means
that these solutions enjoy non-perturbative gravitational stability with
respect to fluctuations of the metric and dilaton. To complete the discussion
we now show that the surface integral (\ref{truth}) form of $E_{WN}$ is well
defined on the boundary of the coordinate chart (\ref{globmet}) in Sec. 2. We specify
the behavior of metric and dilaton perturbations, such that  $E_{WN}$ computes
a finite linear combination of charges of the $AdS_d$ isometry group.

The treatment of Sec. 3 applies with few changes to
$AdS_d$-sliced domain walls. We consider perturbed solutions of
the form (\ref{deform}) with background metric (\ref{globmet}) and accompanying
dilaton. The background frame forms are
\bea \label{fram}
E^{\hat{\mu}} &=& Le^{A(\mu)} d\mu\\
E^{\hat{\lam}} &=&  \frac{Le^{A(\mu)}}{\cos\lam} d\lam\\
E^{\hat{t}} &=&  \frac{Le^{A(\mu)}}{\cos\lam} dt\\
E^{\hat{a}} &=& \frac{Le^{A(\mu)} \sin\lam}{\cos\lam} e^{\hat{a}}
\eea
where $e^{\hat{a}}$ is a frame on $S_{d-2}, ~~a=1,\ldots,d-2$.

The boundary consists of the 3 components shown in Fig. \ref{Jan-pic}b:
\newline
1. The portion at $\mu = -\mu_0$ with $0 < \lam < \frac{\pi}{2}-\delta$
    and volume form
\be \label{vol1}
d\Sigma^{t \mu} = L^2 e^{-2A(\mu)}\cos\lam\, E^{\hat{\lam}}\wedge
E^{\hat{1}}\wedge
\ldots\wedge E^{\widehat{d-2}}
\ee
where $\delta$ is a small positive number. \newline
 2. The keyhole surrounding the corner on which $\lam =
\frac{\pi}{2}-\delta$ and $ -\mu_0 < \mu < \mu_0$ with volume form
\be \label{vol2}
d\Sigma^{t \lam}=- L^2  e^{-2A(\mu)}\cos^2\lam \,
E^{\hat{\mu}}\wedge E^{\hat{1}}\wedge
\ldots\wedge E^{\widehat{d-2}}
\ee
3. The portion $\mu = \mu_0$ with $0 < \lam < \frac{\pi}{2}-\delta$
  and volume form (\ref{vol1}).

An important change is that the Killing spinors to be used in
(\ref{truth}) are those given in (\ref{fake_Killing_sol}) in which
we now replace $\Gamma^{\hat{r}} \rightarrow \Gamma^{\hat{\mu}}$ and define
$\xi = (\frac{\omega'}{\bar{\omega}'})^{\frac{1}{4}}.$
Now let $\Gamma$ denote any matrix of the Dirac
(Clifford) algebra in $d$ dimensions. It is easy to compute the
Killing spinor bilinears
\bea \label{kilbil}
\bar{\eps}_1 \Gamma \eps_2 &=& 2 e^{A} \bar{\xi} \xi \,
\bar{\eps}_{K1}(\Gamma - \Gamma^{\hat{\mu}}\Gamma
\Gamma^{\hat{\mu}})\eps_{K2}\\
\bar{\eps}_1 \Gamma {\bf W} \eps_2 &=& 2
e^{A}\bar{\eps}_{K1}\{\re(\xi^2 \bar{\omega})
[\Gamma^{\hat{\mu}},\Gamma] -\im(\xi^2 \bar{\omega})(\Gamma +  \Gamma^{\hat{\mu}}\Gamma
\Gamma^{\hat{\mu}})\}\eps_{K2}.
\eea
The first equation tells us that $\bar{\eps}_1 \Gamma^\rho \eps_2$ is a
Killing vector of the $(d+1)$-dimensional space-time with
vanishing radial component $( \rho \rightarrow \mu)$. Transverse
components $( \rho \rightarrow i,~~i=0,\ldots,d-1)$   are
proportional to
$e^{A}\bar{\eps}_{K1} \Gamma^i \eps_{K2}$, which is an $AdS_d$
Killing vector, and the full set of these is spanned as we vary
$\eps_{K1},~ \eps_{K2}$.

Let's look first at the last term of (\ref{truth}), which involves the tensor
bilinear $\bar{\eps}_1 \Gamma^{\nu \rho} {\bf W'} \eps_2$. The second equation
in (\ref{kilbil}) applies if we change ${\bf W} \rightarrow {\bf W'}$ on both
sides. The product $\xi^2 \bar{\omega}' =
\sqrt{\bar{\omega}'\omega'}$ is real, so
only the commutator term in (\ref{kilbil}) contributes. On the keyhole part of
the boundary, we find $[\Gamma^{\hat{\mu}}, \Gamma^{t \lam}]$ which vanishes.
On the boundary components at $\mu = \pm \mu_0$, we find $[\Gamma^{\hat{\mu}},
\Gamma^{t \mu}] = -2 E^{\hat\mu \mu} \Gamma^t$. The tensor bilinear thus reduces to a multiple
of the energy Killing vector. Thus the last term of  (\ref{truth}) certainly
vanishes on the keyhole, and we now show that it vanishes on the other two
boundary components by examining the behavior of the integrand as $\mu
\rightarrow \pm \mu_0$. We note the behavior $\sqrt{{\bf W'}^2} \sim
\bar{\phi}'\sim e^{-dA(\mu)}$, which follows from (\ref{phiflow}) and the
property of dilaton in the solution of \cite{Gutp} noted above our (2.16).
Using (\ref{fram}, \ref{kilbil}), we find that the factor $\bar{\eps}_1
\Gamma_{t \mu} {\bf W'} \eps_2 d\Sigma^{t \mu}$ is constant on the boundary.
However the normalizable dilaton fluctuation vanishes on the boundary at the
rate $\varphi \sim (\mu \mp \mu_0)^d \sim e^{-2A(\mu)}$. Thus the last term of
(\ref{truth}) vanishes for our dilaton domain walls.

Let's look next at the terms of (\ref{truth}) involving $\bar{\eps}_1
\Gamma^{\rho \sigma} {\bf W} \eps_2~d\Sigma_{\nu \tau}$ with various index
assignments. On the boundary components $\mu = \pm \mu_0$, the product $\xi^2
\bar{\omega}$ is real, as follows from (\ref{nofake3}) or (\ref{phib}).
Thus
only the commutator term contributes in (\ref{kilbil}) and it is non-vanishing
for index combinations $\Gamma^{\mu i}$ only. It then follows from
(\ref{fram},~\ref{kilbil}) that $\bar{\eps}_1 \Gamma^{\mu i} {\bf W} \eps_2$
vanishes as $e^{-A(\mu)}$, and is proportional to an $AdS_d$ Killing vector.
Clearly, $g^{\rho\sigma}~\sim~ e^{-2A(\mu)}$. The volume element behaves as
$d\Sigma_{t\mu}~\sim~e^{(d+1)A}$, while normalizable metric fluctuations vanish
at the rate $h_{\rho\sigma}~\sim~ e^{(2-d)A(\mu)}.$ Putting these factors
together, we see that the terms under consideration give a finite contribution
to the energy of a deformed domain wall.

To analyze the behavior of the tensor bilinear terms on keyhole, we must take
the limit $\delta \rightarrow 0$, which is the boundary limit $\cos({\lam})
\rightarrow 0$ on the $AdS_d$ slices. We discuss this limit first for the bulk
space-time $AdS_{d+1}$ with $AdS_d$ slicing and then adapt the argument to the
dilaton domain wall geometry.

In Sec. 2 of \cite{Gutp}, global metrics for $AdS_{d+1}$ with
both standard and $AdS_d$ slicing are both derived from the
embedded hyperboloid description: $X_0^2 +X_{d+1}^2 -X_1^2- \cdots
X_d^2 =L^2.$ The two metrics are
\bea
ds^2 &=& \frac{L^2}{\cos^2\theta} \l( -dt^2 + d\theta^2 +
\sin^2\theta\,  d\Omega^2_{d-1}\r) \nonumber \\
     &=& \frac{L^2}{\cos^2\mu~\cos^2\lam}\l(-dt^2 + \cos^2\lam \,
d\mu^2 +d\lam^2 + \sin^2\lam\, d\Omega^2_{d-2}\r)
\eea
Comparison of the conformal factors yields one relation between the two sets of
coordinates, namely $\cos\theta ~=~\cos\mu \cos\lam.$ A normalizable mode of a
scalar field transforming in a representation of the isometry group $SO(d,2)$
with lowest weight $\Delta$ of the $SO(2)$ generator (the energy) vanishes at
the rate $(\cos\theta)^{\Delta}$ on the $AdS_{d+1}$ boundary. When expressed in
terms of the coordinates for $AdS_d$ slicing it therefore vanishes at the rate
$(\cos\lam)^{\Delta}$ as $\lam \rightarrow \frac{\pi}{2}.$ For the massless
dilaton $\Delta=d$. We need the corresponding result for metric fluctuations $h_{\mu\nu}$.
In the ``axial gauge'' $h_{\mu \mu}=h_{\mu i}=0$, $\,h_{ij}$ is related
by $h_{ij} = e^{2A} \tilde{h}_{ij}$ to the field $\tilde{h}_{ij}$, whose 
wave equation is the same as that of a massless scalar.
Thus normalizable modes of $\tilde{h}_{ij} ~\sim~ (\cos\lam)^d$.

We use this rate to obtain the behavior of the tensor terms of (\ref{truth}) as
the keyhole boundary contribution shrinks toward the corner. We need the fact
that the $AdS_d$ Killing spinors behave as
$\eps_K~\sim~(\cos\lam)^{\frac{1}{2}}$, and that the volume element behaves
as $d\Sigma_{t \lam}~\sim~ (\cos\lam)^{-d}$. It is convenient to work in the
axial gauge. Detailed inspection of the various tensor components in
(\ref{truth}) shows that they vanish at least as fast as $ (\cos\lam)^3$.  The
analysis so far is valid for $AdS_{d+1}$. However, the domain wall space-time
shares the isometry $SO(d-1,2)$ and may be viewed as a small distortion of
$AdS_{d+1}$ when the parameter $b$ of (\ref{profeq}) is small. Therefore we
expect at most a small modification of the exponent in the behavior
$h_{ij}~\sim~(\cos\lam)^d$ we assumed. Thus we reach the conclusion that the
contribution of tensor terms on the keyhole part of the boundary vanishes as
$\delta \rightarrow 0$.

It is now straightforward to analyze the boundary behavior of the terms in
(\ref{truth}) involving the Killing vector bilinears. Using the asymptotics of
the metric fluctuations $h_{ij}$ discussed above, we find a vanishing
contribution from the keyhole at the rate $(\cos\lam)^3$ as $\delta \rightarrow
0$ and a finite contribution from the boundary components at $\mu = \pm \mu_0.$

In summary, we have shown that $E_{WN}$ computes a linear
combination of the $AdS_d$ charges for any deformation of the
dilaton domain wall metric solution which satisfies the asymptotic
conditions stated above. The energy of such a deformation is
positive. The keyhole part of the boundary does not contribute.

\section{Stability with additional scalar fields.}
\setcounter{equation}{0}

The stability argument developed in Sec. 4 strictly applies to models with
action (\ref{toymod}) containing only a single scalar field. At the formal
level it is straightforward to add additional scalars, but the equations
(\ref{VWads},~\ref{plane} -- \ref{def_alpha}) which determine the
superpotential {\bf W} become partial differential equations in field space,
and it is more difficult to show that {\bf W} exists. However, it is important
to extend our results for the stability of the Janus solution of Type IIB
supergravity to include the additional fields which appear in compactifications
to 5 dimensions. In this section we develop a reasonably general stability
criterion, related to the approach of \cite{Boucher}. We then test this
criterion in several known consistent truncations of Type IIB supergravity
which involve the negative $m^2$ scalars with potentials unbounded below. These
fields are certainly the main threat to stability, and it is gratifying that the
test is satisfied in all cases examined.

The new criterion applies to dilaton domain walls in theories containing
the dilaton $\phi$ plus additional scalars $\psi^a$ with action
\be \label{newmod}
S =  \int d^{d+1}x \sqrt{-g} \l[ \frac{1}{2\kappa^2} R -
\frac{1}{2} \partial_\mu \phi \partial^\mu \phi
- \frac{1}{2} \partial_\mu \psi^a \partial^\mu \psi^a - V(\psi^a)\r].
\ee
We assume that the potential $V(\psi^a)$ does not depend on the dilaton, and
that there is a scalar superpotential $U(\psi^a)$ which is related to $V$ by
(with $U,_a \equiv \frac {\partial U} {\partial \psi^a}$)
\be \label{VUrel}
V = p U,_aU,_a - q U^2.
\ee
In our conventions, the constants are given by
\bea \label{conv}
p &=& \frac{2(d-1)^2}{\kappa^4}\\
q &=& \frac{2d(d-1)}{\kappa^2}
\eea
as in (\ref{VWrel}), but we allow different values to facilitate
comparison with models in the literature which use different
conventions, but $p,q >0$ always.
In the models we study below $U(\psi^a)$ is a true supergravity
superpotential generated in the truncation from 10 to 5
dimensions, but it could also be a fake supergravity
superpotential obtained as a solution to ({\ref{VUrel}) viewed as
a partial differential equation for $U(\psi^a)$.
We also assume that
\bea
U,_a|_{\psi^b =0} &=& 0\\
V_0 \equiv  V(0) &=& -\frac{d(d-1)}{2L^2 \kappa^2}
\eea
so that the equations of motion of the enlarged system have the same
$AdS_d$-sliced dilaton domain wall solution discussed in Sec. 2 with all
$\psi^a=0$. We let ${\bf W}(\phi)$ denote the superpotential obtained in
Sec. 4 for the dilaton domain wall.

Our strategy \cite{Boucher} is to find a new superpotential ${\cal
W}(\phi,\psi^a)$ to be inserted in the covariant derivative
(\ref{3.cov}) of the Witten-Nester integral. The new form should
have the property that the last term in (\ref{hotstuff}) is
replaced by
\be
p({\cal W},_\phi)^2 + p{\cal W},_\a{\cal W},_\a -q{\cal W}^2
- V(\psi^a) \le 0.
\ee
The last term of (\ref{hotstuff}) will not vanish in general as it does for a
true adapted superpotential, but it is non-negative. At the critical point
$\psi^a=0$ it will vanish, thus guaranteeing stability.

It is quite straightforward to show that the empirically inspired
form
\footnote{We take the explicit matrix square root as 
$ {\cal W} \equiv \W \sqrt{1 +\frac{1}{\W^2} \l( U(\psi^a)^2
+\frac{V_0}{q}\r)} $.}
\be \label{newsup}
{\cal W}(\phi,\psi^a) \equiv \sqrt{ {\bf W}(\phi)^2 +U(\psi^a)^2
+\frac{V_0}{q}}
\ee
satisfies
\be \label{ident}
p\l[({\cal W},_\phi)^2 + {\cal W},_\a{\cal W},_\a\r] -q{\cal W}^2 - V(\psi^a) =
-\frac{p {\bf W}'^2}{q {\cal W}^2}\l(p U,_aU,_a + q U^2 + V_0\r).
\ee
Thus non-perturbative stability will hold if
\be \label{ineq}
p U,_aU,_a + q U^2 + V_0 \ge 0.
\ee
Furthermore, since ${\cal W}(\phi,0) \equiv {\bf
W}(\phi)$, the energy of the dilaton domain wall background,
evaluated using $E_{WN}$ with the new $\hat{\nabla}$ operator,
vanishes, and this background has the same $AdS_d$ Killing
spinors found in Sec. 4.

As we will see below, the inequality (\ref{ineq}) is not a general  property of
superpotentials in supergravity. However, it is quite simple to check that it
is valid in several known consistent truncations of Type IIB supergravity which
involve scalars of negative $m^2$ and potentials unbounded below.

The simplest model contains a single scalar whose mass, namely $m^2=-4$,
saturates the BF bound. It is a special case \cite{Hertog} of more general
models \cite{fgpw2} considered in the framework of gauged ${\cal N}=8$
supergravity \cite{grw,ppvn}. With $\kappa^2 L^2 =1$, the potential is
\be
V(\psi) = -  2 e^{\frac{2\psi}{\sqrt{3}}}
 - 4e^{-\frac{\psi}{\sqrt{3}}},
\ee
and one easily finds the superpotential (using (\ref{conv}))
\be
U(\psi)= \frac{1}{3}e^{\frac{\psi}{\sqrt{3}}} + \frac{1}{6} e^{-\frac{2
\psi}{\sqrt{3}}}.
\ee
One can check directly that (\ref{ineq}) is satisfied.

The general model of this type \cite{fgpw2} involves 5 independent scalars with
$m^2=-4$. The potentials is a sum of exponentials of linear combinations of
these fields. A special case involving 2 non-vanishing scalars was also derived
from the viewpoint of consistent truncations of the Type IIB theory in
\cite{Cvetic:1999xp}. The analysis of these models is somewhat more involved,
but one can also show that (\ref{ineq}) is satisfied. Since the left hand side
of (\ref{ineq}) is bounded, it is enough to check the inequality for the local
minima and at infinity. Given the explicit form of the superpotential $U$ one
can easily show that the matrix $p\, U,_{ab} + q\, \delta_{ab}U$ is strictly
positive definite and hence all the minima are zeros of $U,_{a}$ which greatly
simplifies the analysis.

A different subtheory of gauged ${\cal N}=8$ supergravity
with potential unbounded below contains \cite{fgpw1}
scalars with masses $m^2 = -4$ and $m^2=-3$. The simplest version contains 2
fields, called $\psi_1,\psi_3$ and the superpotential
\be
U \sim \frac{1}{4 \rho^2}\l[ \cosh(2\psi_1)(\rho^6-2) -3\rho^6-2\r]
\ee
and $\rho = \exp\l(\frac{\psi_3}{\sqrt{6}}\r)$. Using the conventions of
\cite{fgpw1}, one also finds that (\ref{ineq}) holds. A more general
version with 3 negative $m^2$ scalars was studied numerically. Again
(\ref{ineq}) is valid.

There do not appear to be any consistent truncations of Type IIB supergravity
which involve both positive and negative $m^2$ scalars, but several involve
only positive $m^2$ fields. the simplest of these \cite{Bremer:1998zp} contains
the dilaton $\phi$ and the breathing mode $\psi$ with $m^2_{\psi} = 32$. The
potential, which is bounded below, and superpotential are
\bea
V(\psi) &=& \frac{1}{\kappa^2 L^2} \l[4 e^{8 \alpha \psi} - 10
e^{\frac{16\alpha}{5} \psi} \r]
\nonumber\\
U(\psi) &=& \frac{1}{3 L} \l[ e^{4 \alpha \psi} -\frac{5}{2}
e^{\frac{8\alpha}{5} \psi}
\r]
\\\nonumber
\alpha &=& \frac{1}{2}\sqrt{\frac{5}{6}} \, \kappa \,.
\eea

It is easy to see that in this case the inequality (\ref{ineq}) is violated for
large negative $\psi$. However the superpotential ${\cal W}(\phi,\psi)$ which
provided the appropriate bound for truncations with negative $m^2$ need not
work universally. For the breathing mode model, we can simply take the matrix
superpotential ${\bf W}(\phi)$ of Sec. 4. The quantity
\be
p {\bf W}'^2 -q {\bf W}^2 - V(\psi)
\ee
which appears in (\ref{hotstuff}) is negative for all nonzero $\psi$, which is
sufficient to establish stability.

It is curious to note that another simple candidate superpotential, namely the
product ${\cal W} \equiv  {\bf W}(\phi) U(\psi^a)/U(0)$, produces the
inequality
\be
p({\cal W},_\phi)^2 + p{\cal W},_\a{\cal W},_\a -q{\cal W}^2
- V(\psi^a) = p  {\cal W}'^2 U^2 \ge 0
\ee
of the wrong sense for stability in all the models above.

Further improvements of the arguments above may well be
possible. However, we shall be content for the present with the
non-perturbative stability arguments presented for the Janus solution
which involve fluctuations of the metric, the dilaton, and
several examples of negative $m^2$ scalars.

\section{A curious linear dilaton solution}
\setcounter{equation}{0}
\label{Section_LinearDilaton}

In (\ref{lindil}) of Sec. 4, it was noted that for
constant potential $V(\phi) =V_0$ of (\ref{adssol}), there is a simple
$su(2)$ superpotential
\be
\W(\phi)= \frac{1}{2L}\sqrt{\frac{d}{d-1}}
\l(\begin{array}{cc} 0 & \bar\zeta(\phi) \\
                      \zeta(\phi) & 0
    \end{array}\r).
\ee
\be
\zeta(\phi) \equiv \exp\l(i\kappa\frac{\phi}{\sqrt{d-1}}\r)
\ee
which appears among more complicated implicit solutions. As a simple
consistency check of our formalism we now find the solution $\phi(r),~A(r)$ of
the first order flow equations (\ref{phiflow},~\ref{Aalg}) for this $\W(\phi)$
and show that it is a solution of the second order equations of motion
(\ref{adseom}) or, equivalently, (\ref{geneom}).

First we compute $\W'(\phi)$, note that
$\{\W(\phi),\W'(\phi)\}=0$, and that the invariants
\bea \label{invs}
\W^2 &=& \frac{d}{4L^2(d-1)}\\
\W'^2 &=& \frac{\kappa^2 d}{4L^2(d-1)^2}
\eea
are correctly related to the potential by (\ref{VWads}).

The flow equation (\ref{phiflow}) gives the solution
\be
\phi(r) = -\frac{\sqrt{d}}{\kappa L} (r-r_0)
\ee
The compatibility condition (\ref{intcon2}) implies that $A'=0$, and
(\ref{Aalg}) then gives (for $L_d= L$)
\be
e^{2A} = \frac{d-1}{d}.
\ee
The linear scalar obviously satisfies the scalar equation of
(\ref{adseom}), and it is easy to check that the second equation
in (\ref{adseom}) is also satisfied.

The line element (\ref{adswall}) of this solution is
\be
ds^2 = \frac{d-1}{d} \bar{g}_{ij}(x) dx^i dx^j + dr^2,
\ee
where $ \bar{g}_{ij}(x)$ is an $AdS_d$ metric.
Thus we find the non-singular geometry $AdS_d \otimes {\cal R}$ with
accompanying linear scalar. One can verify directly that
(\ref{geneom}) is satisfied.\footnote{This solution was found previously 
in \cite{Robb} where linearized stability analysis was performed. The solution was 
also found in \cite{Quevedo}. We thank A. Kehagias for pointing this out to 
us.} 
It would be interesting to study the stability
of this solution whose boundary structure differs from that considered 
in previous sections.

For $d=4$ this solution can be lifted to Type IIB by adjoining an
$S_5$ and self-dual 5-form. The full system is
\bea
ds_{10}^2 &=&  \frac{3}{4} \bar{g}_{ij}(x) dx^i dx^j
+ dr^2 + l^2 d\Omega_5^2\\
\phi(r) &=& - \frac{2}{\kappa L} (r-r_0)\\
F_{\a\b\g\d\e} &=& s_0 ~\eps_{\a\b\g\d\e}
\eea
where $ d\Omega_5^2$ is the metric on the unit 5-sphere, and $\a\b\g\d\e$ are
5-sphere coordinates. We require that this satisfy the 10-dimensional equations
of motion
\be
\frac{1}{\k^2} R_{MN} = \partial_M \phi \partial_N \phi +
\frac{1}{96} F_{MPQRS} F_N^{PQRS},
\ee
which quickly gives the scales $l=L$ and $s_0 = 4 L^4/\kappa$.
Until stability is established, it is premature to speculate about
a possible physical application of this simple non-singular solution 
of Type IIB supergravity.

\section*{Acknowledgments}

We thank Ofer Aharony, Neil Constable, Oliver deWolfe, Andreas Karch, David Tong and 
especially Gary Gibbons for useful discussions.
DZF, CN and MS acknowledge the hospitality of the Benasque Center for Physics, 
where part of this work was done.
The research of DZF is supported by the National Science Foundation Grant PHY-00- 96515, 
and  DZF, CN and MS are supported in part by DOE contract \#DE-FC02-94ER40818. 
KS is supported by NWO.

\newpage

\appendix

\section{Connection 1-forms and curvature tensor for domain walls}
\setcounter{equation}{0}

Let us start with flat domain walls in $d+1$ dimensions with the metric ansatz
\be
ds^2= e^{2A(r)} \eta_{ij} dx^i dx^j + e^{2h(r)}dr^2.
\ee
We introduce the vielbeins
\be
E^{\hat i}= e^{A}dx^i, \qquad  E^{\hat r}= e^{h}dr.
\ee
The hat over an index indicates that it is a frame index. The range of indices
$i$ and $j$ will be always taken $0, \ldots d-1$.
The spin connection one forms are given by
\be
\omega^{\hat i \hat r} = A' e^{-h} E^{\hat i}, \qquad  \omega^{\hat i \hat j} = 0.
\ee
Nonzero components of the  Ricci tensor (in curved indices) are then
\bea
R_{ij} &=& -\eta_{ij} (d A'^2 + A'' - A' h' )  e^{2 A - 2 h},
\nonumber\\
R_{rr} &=& -d ( A'' + A'^2 - A'h').
\eea

Now, let us consider $AdS_d$-sliced domain walls with the metric
\be
ds^2= e^{2A(r)} \bar{g}_{ij} dx^i dx^j  + e^{2h(r)}dr^2,
\ee
where $\bar{g}_{ij}$ is a metric on the $AdS_d$ slices.
In this case our choice of vielbeins is
\be
E^{\hat i}= e^{A} \bar{e}^{\hat i}, \qquad  E^{\hat r}= e^{h}dr.
\ee
where we have denoted with $\bar{e}^{\hat i}$ the vielbein for $AdS_d$.
The spin connection is now
\be
\omega^{\hat i \hat r} = A' e^{-h} E^{\hat i}, \qquad
\omega^{\hat i \hat j} = {\bar\omega}^{\hat i \hat j},
\ee
where ${\bar\omega}^{\hat i \hat j}$ is the spin connection on the $AdS_d$ slices,
whose explicit form is not needed.                    .
Nonzero components of the  Ricci tensor are given by
\bea
R_{ij} &=& {\bar R}_{ij}- \bar{g}_{ij} (d A'^2 + A'' - A' h' )  e^{2 A - 2 h},
\nonumber\\
R_{rr} &=& -d ( A'' + A'^2 - A'h'),
\eea
where
\be
{\bar R}_{ij} =  -\frac{d-1}{L_d^2} g_{ij}
\ee
is the Ricci tensor of $AdS_d$ space of scale $L_d$.

\section{Further information on Janus domain walls}
\setcounter{equation}{0} \label{App-d2}

\subsection{Explicit form of solution for $d=2.$}

The metric for $AdS_2$ sliced domain walls in $AdS_3$ in the $r$-coordinate
takes the form
\be
ds^2 = e^{2A(r)} ds_{AdS_2}^2 + dr^2.
\ee
The explicit solution of the equations of motion is
\bea
A(r) &=& \frac{1}{2} \log \l(\frac{1}{2} \l(1+\sqrt{1-4b} \cosh 2r \r) \r)
\nonumber\\
\phi(r) &=& \phi_0 + \frac{\sqrt{2}}{\kappa} \arctanh
\frac{(1-\sqrt{1-4b})\tanh r}{2\sqrt{b}}.
\eea
This is the solution for $L=L_d=1$. To restore dependence on the
scale $L$, one just  replaces $r$ by $r/L$. The relation of the
constant $b$ to $c$ defined by $\phi'=c e^{-d A}$ is
\be
b =  \frac{c^2 \kappa^2 L^2}{d(d-1)}.
\ee
Two coordinate independent features are evident. First the critical value of
$b$ beyond which the geometry contains a naked singularity is $b=\frac{1}{4}$.
Second the asymptotic values of $\phi$ on the two components of the boundary
are
\be
\phi_{\pm \infty} = \phi_0 \pm \frac{\arctanh 2\sqrt{b}}{\sqrt{2}\, \kappa}.
\ee

\subsection{Radial coordinate $\mu$. }

After change of variable, the integral (\ref{muint}) which
defines the wall profile can be written as
\be\label{muA}
\mu
=  \int_x^{x_{min}} dx \frac{1}{\sqrt{1-x^2+b\, x^{2d}}},
\ee
where $x_{min}$ is the smallest positive root of the polynomial
in the denominator. The maximum value of $\mu$ is
\be\label{mumax}
\mu_{0} = \int_0^{x_{min}} dx \frac{1}{\sqrt{1-x^2+b\, x^{2d}}}.
\ee
Series expansion in the parameter $b$ gives
\be
x_{min} = 1 + \frac{1}{2} b + \frac{4d-1}{8} b^2 + O(b^3).
\ee
Calculating the expansion of $\mu_0$ to order $b^2$ and arbitrary $d$ it is
easy to guess the form of the expansion to all orders in $b$
\be\label{mumax-exp}
\mu_{0} = \frac{\pi}{2} \sum_{n=0}^\infty \frac{b^n}{n!} \,
\frac{\Gamma\!\l(nd+\frac{1}{2}\r)}{\Gamma(n(d-1)+1)\,
\Gamma\!\l(\frac{1}{2}\r)}.
\ee
We have verified this formula to all orders in $b$ analytically for $d=1,2$ and
numerically for $d=4$. The convenient form
\be \label{convenient}
\mu_{0}-\mu = \int_0^{x} dx \frac{1}{\sqrt{1-x^2+b\, x^{2d}}}
\ee
yields the series expansion of $\mu$ in terms of $x=e^{-A}$
\be\label{mu_expansion}
\mu_{0}-\mu = \arcsin x - b\, \frac{x^{2d+1}}{2(2d+1)}\,
{_{2}F_{1}}\l(d+\frac{1}{2},\frac{3}{2},d+\frac{3}{2},x^2\r) + O(b^2 x^{4d+1}).
\ee
Inverting the series we find
\bea\label{x_expansion}
&& e^{-A(\mu)} \equiv x = \sin(\mu_{0}-\mu)
\nonumber\\
&& \qquad + \frac{b}{2(2d+1)} \sin^{2d+1}(\mu_{0}-\mu) \cos(\mu_{0}-\mu)\,
{_{2}F_{1}}\l(d+\frac{1}{2},\frac{3}{2},d+\frac{3}{2},\sin^2(\mu_{0}-\mu)\r)
\nonumber\\
&&\qquad + O(b^2 \sin^{4d+2}(\mu_{0}-\mu)).
\eea
Near the boundary  $\mu \approx \mu_{0}$ the form of the scale factor is
\be\label{warp-as}
e^{2A(\mu)} \approx \frac{1}{\sin^2 \l(\mu - \mu_{0} \r)}\l[1 + O(\mu-\mu_0)^{2d}\r].
\ee

The equations above define $A(\mu)$ in the region $0 \le \mu <
\mu_0$. However, as discussed in Sec. 2, it can be extended as an
even $C^{\infty}$ function to the full range $-\mu_0 < \mu <\mu_0$.

In the special case $d=2$ we can integrate (\ref{convenient}) and
invert to obtain the explicit solution
\be\label{xex}
e^{-A(\mu)} \equiv x = \gamma \sn\l(\frac{1}{\gamma} (\mu_0-\mu), \sqrt{b}
\gamma^2 \r),
\ee
where
\be
\gamma = x_{min} = \frac{\sqrt{2}}{\sqrt{1+\sqrt{1-4b}}}
\ee
is the smallest positive root of the equation $1-x^2+b x^{4} = 0$ and
$\sn(u,k)$ is the standard Jacobi elliptic function. Note that the metric is
{\it doubly periodic}\footnote{Real periodicity in $\mu$ can be proved to exist
for all dimensions. The second complex period is special to $d=2$ and it would
be interesting to see if it has a deeper meaning.} in the coordinate $\mu$. The
real period is
\be
4\gamma K(\sqrt{b} \gamma^2).
\ee
One may easily check using the definition of the complete elliptic
integral that this is the same as $4\mu_{0}$. The period clearly
blows up as $b$
approaches its critical value $\frac{1}{4}$ which corresponds to
$\gamma=\sqrt{2}$.

\end{document}